\documentclass[prb,twocolumn, amsmath,amssymb]{revtex4}

\setcitestyle{numbers,square}
\usepackage{amsmath,amssymb}
\usepackage{tabularx}
\usepackage{graphicx}
\usepackage{bmpsize}
\usepackage{bm}
\usepackage{color}
\usepackage{amssymb}
\usepackage[version=3]{mhchem}
\DeclareMathAlphabet{\mathpzc}{OT1}{pzc}{m}{it}
\newcommand{\nn}{\nonumber}

\begin{document}
\makeatletter
\renewcommand\@biblabel[1]{[#1]}
\makeatother

\preprint{APS/123-QED}

\title{Electronic properties of graphyne-$N$ monolayer and its multilayer:  \\
even-odd effect and topological nodal line semimetalic phases}

\author{Takuto Kawakami}
\email{t.kawakami@qp.phys.sci.osaka-u.ac.jp}
\affiliation{Department of Physics, Osaka University, Toyonaka, Osaka 560-0043, Japan}
\author{Takafumi Nomura}
\affiliation{Department of Physics, Osaka University, Toyonaka, Osaka 560-0043, Japan}
\author{Mikito Koshino}
\affiliation{Department of Physics, Osaka University, Toyonaka, Osaka 560-0043, Japan}

\date{\today}

\begin{abstract}
We study the electronic structure and topological properties of monolayer and ABC-stacked multilayer of graphyne-$N$,
which are a family of planar carbon sheets consisting of $sp$ and $sp_2$-bonding. 
By using the density-functional theory and the effective continuum model,
we find a striking even-odd effect in the dependence of the band structure on $N$ (the number of carbon-carbon triple bonds between neighboring benzene rings).
Specifically, even-$N$ graphyne monolayer has doubly-degenerate conduction and valence bands near the Fermi energy,
and in its ABC multilayer, the band inversion of the doubly-degenerate bands leads to a nodal-line semimetal phase
with non-trivial $\mathbb{Z}_2$ monopole charge.
In contrast, odd-$N$ monolayer has singly-degenerate bands in separate valleys,
and its ABC multilayer can have only $\mathbb{Z}_2$-trivial nodal lines.
ABC graphynes with larger $N$ tend to be trivial insulators because of smaller interlayer coupling,
while the external pressure induces a topological phase transition from the trivial phase to the nodal line semimetal phase.
\end{abstract}


\maketitle
\section{Introduction}

Carbon allotropes have attracted much attention because of their variety of topological properties depending on the atomic configurations.
The best known example is graphene, where a honeycomb lattice of carbon atoms constructed by $sp_2$ covalent bonds~\cite{netoreview}
gives rise to a symmetry-protected band touching with a linear dispersion~\cite{wallace1947, slonczewski1958}.
The $sp$-based one-dimensional (1D) carbon chain, carbyne~\cite{smith1982},  
is known to be a 1D topological insulator described by Su-Schriefer-Heeger model~\cite{su1979, rice1983}.
Other types of pure carbon materials have also been studied from a viewpoint of topological phenomena~\cite{koshino2013, ju2015, yin2016, slizovskiy2019,cao2017,rizzo2018,tamaki2020,izumida2016,chen2015,ruegg2013, po2018, ahn2019prx, koshino2019, moon2012,song2019, park2019, chenreview, bouhon2019}.

\begin{figure}[b]
\begin{center}
	\includegraphics[width=85mm]{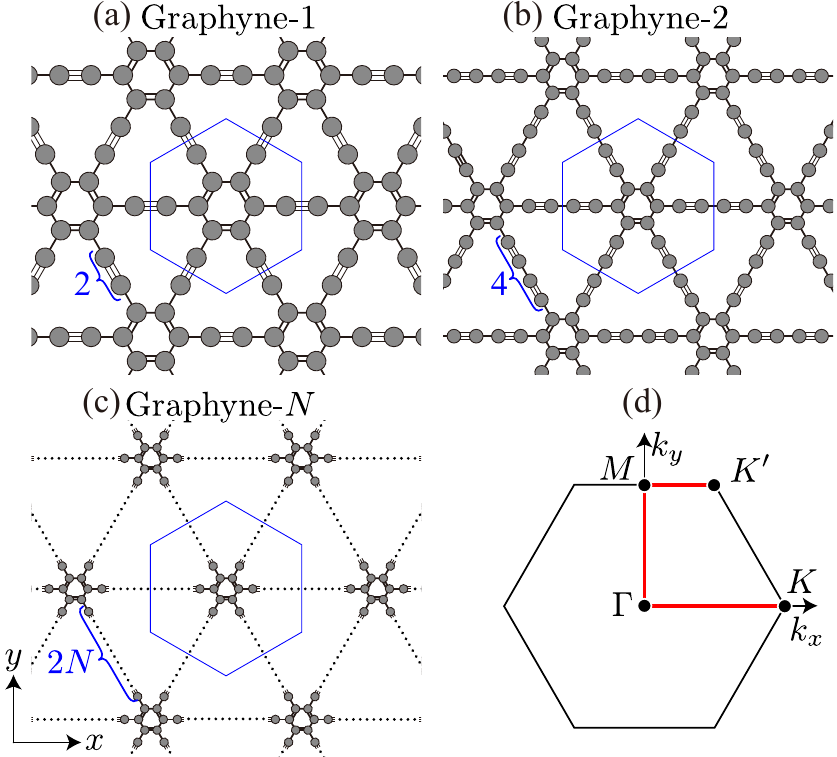}
	\caption{(a) Lattice structures of monolayer graphyne-1, (b) -2, and (c) -$N$,
	with a unit cell indicated by a blue hexagon.
	(d) Brillouin zone and the high-symmetry line along which the band calculation is performed.
	}\label{fig:lattice}
\end{center}
\end{figure}

Furthermore, there is another class of carbon allotropes called graphyne, 
which can be regarded as a hybrid of graphene and carbyne.
A graphyne generally takes a two-dimensional planar structure purely composed of carbon atoms
with a mixture of $sp_2$ and $sp$ bonds~\cite{baughman1987,peng2014,narita1998,narita2000, malko2012, puigdollers2016, kim2016}. 
Figure \ref{fig:lattice} shows representative structures called the graphyne-$N$ family,
which is defined as a triangle lattice of benzene rings connected by polyyne chains with $N$ triple bonds (\ce{C#C}).
The band calculation showed that monolayer graphyne-$N$'s are semiconductors with band gaps about a few hundred meV \cite{narita1998, peng2014}. 

Recently, a significant progress has been made on the experimental studies on graphyne-2 (or called graphdiyne)\cite{gli2010,matsuoka2017}.
Particularly, a high-quality three-dimensional stack of graphyne-2 was successfully synthesized,
and it was identified as ABC rhombohedral structure by X-ray defraction measurement~\cite{matsuoka2017}.
Theoretically, two of the present authors found that ABC-stacked graphyne-2 is 
a topological nodal line semimetal with band-touching nodal loops,
and derived an effective continuum model for its low-energy band structure~\cite{nomura2018}.
Then it was pointed out that the nodal line of ABC-graphyne-2 is characterized by a nontrivial $\mathbb{Z}_2$ monopole charge~\cite{ahn2019}.
A $\mathbb{Z}_2$-nontrivial nodal loop cannot disappear in its own, but it can only pair-annihilate with the other loop\cite{fang2015, fangreview}.
In recent years, various materials with nodal line are theoretically proposed and some of them are experimentally probed~\cite{ 
fangreview, burkov2011, lu2013,  ychen2015, hskim2015, weng2015, mullen2015, ykim2015, yu2015, heikkila2015, 
yamakage2016,bian2016, wang2016, neupane2016, schoop2016, li2016,  koshino2016, lian2016, ezawa2016, huang2016, 
hyart2016, zhu2016, takane2016, bzdusek2016, zhao2016, xu2017, du2017, zhang2017, hirayama2017, kawakami2017,
chen2017, yan2017, ezawa2017, pychang2017, li2017, gchang2017, bi2017, feng2018, takane2018, gao2018, gong2018, 
gavrilenko2018, zhou2018, yan2018, hirayama2018, nakamura2019, jeon2019, li2019, wang2019, ezawa2019, hli2019, 
zhou2020}.
Nevertheless, to the best of our knowledge, nontrivial $\mathbb{Z}_2$ monopole charge~\cite{fang2015} is still elusive in nature.
So far it is known only in ABC-stacked graphyne-2~\cite{ahn2019} and transition metal dichalcogenide~\cite{wang2019,ezawa2019}.

These non-trivial properties of graphyne-2 motivates us to generalize the theoretical analysis to other types of graphyne-$N$.
In this paper, we study the electronic structure and the topological property of general graphyne-$N$ monolayer and its three-dimensional ABC stack,
by using the density functional theory (DFT) and the effective continuum model.
We demonstrate that the band structure is significantly different between even $N$'s and odd $N$'s.
First, we show that even-$N$ graphyne monolayer has doubly-degenerate conduction and valence bands near the Fermi energy,
while odd-$N$ graphyne monolayer only has singly-degenerate bands in separate valleys.
Such the even-odd effect is crucial in the topological nature in ABC-stacked multilayers.
In ABC even-$N$ graphynes, the band inversion of doubly-degenerate bands in monolayer 
leads to a  $\mathbb{Z}_2$-nontrivial nodal-line semimetal phase.
In contrast, the ABC odd-$N$ graphynes can only have $\mathbb{Z}_2$-trivial nodal lines. 
The ABC graphynes with $N \geq 3$ become trivial insulators because of smaller interlayer coupling,
while we demonstrate that the external pressure induces band inversion and topological phase transition to the nodal line semimetal phase.

This paper is organized as follows.
In Sec.~\ref{sec:single}, we study the band structure of monolayer graphyne-$N$ systematically 
in terms of DFT calculations and tight-binding model, and describe the even-odd effect of  band structure from the symmetry consideration.
We investigate the band structures of ABC-stacked graphynes for even and odd $N$ in Sec.~\ref{sec:abceven} and \ref{sec:abcodd},
respectively, where the emergence of nodal line semimetal phase is argued from the symmetrical point of view. 
Finally, the brief summary is presented in Sec.~\ref{sec:summary}.
\begin{figure}[b]
\begin{center}
	\includegraphics[width=85mm]{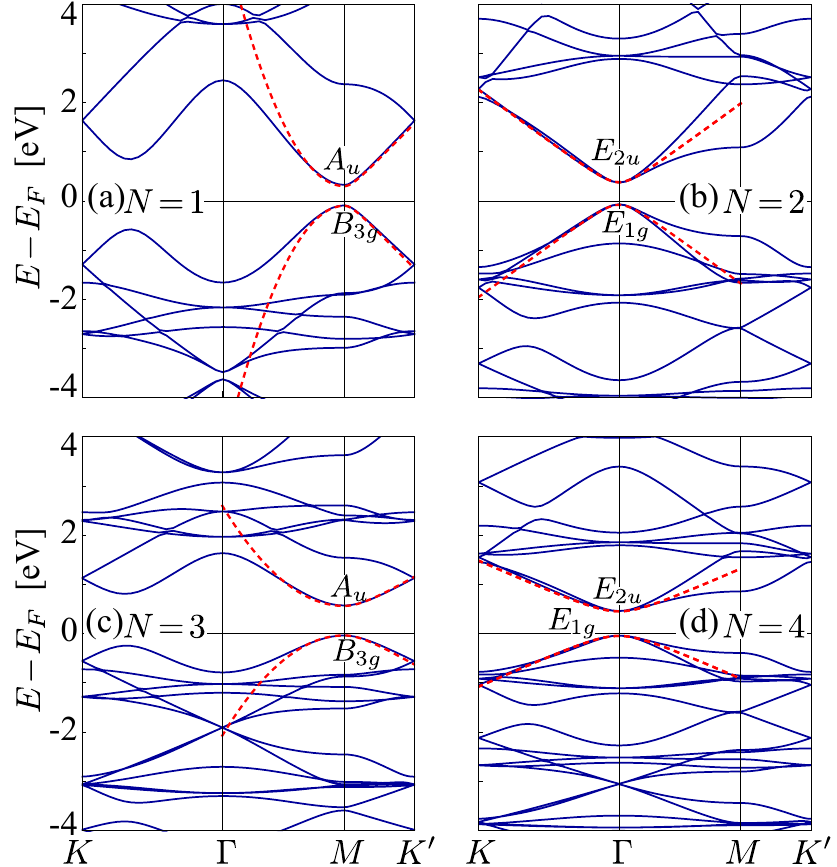}
	\caption{Band structures of monolayer graphyne of $N\!=\!1$, 2, 3, and 4
	obtained from the DFT calculation. 
	The dashed curve is the dispersion relation calculated by the effective continuum model
	(see the text).
	}\label{fig:band2d}
\end{center}
\end{figure}

\begin{figure}[b]
\begin{center}
	\includegraphics[width=85mm]{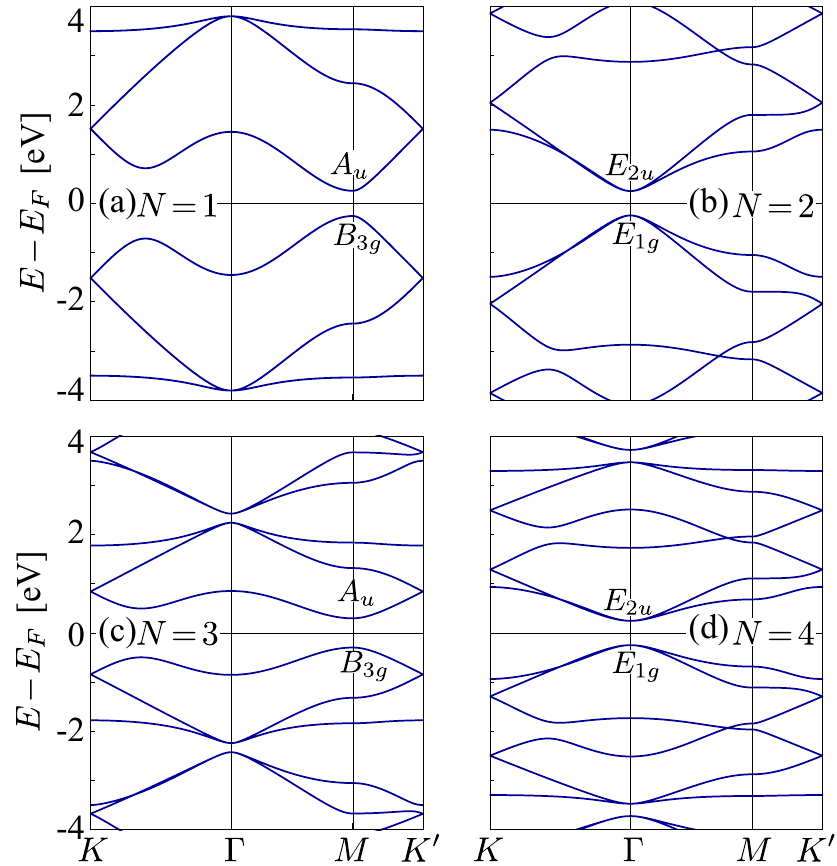}
	\caption{ 
	Band structures of monolayer graphyne $N\!=\!1$, 2, 3, and 4
	obtained from the nearest neighbor tight-binding model. 
	Hopping parameters are determined by Eq.~(\ref{eq:hop2d}) with 
	atomic distances obtained by first-principles calculation.
	}\label{fig:tb2d}
\end{center}
\end{figure}

\section{Monolayer graphynes}\label{sec:single}

\subsection{Band structure}\label{sec:band2d}

We first examine the band structures and the symmetrical properties of monolayer graphyne-$N$.
The lattice structures of monolayer graphyne-1, 2, $\cdots N$ are displayed in Fig.~\ref{fig:lattice}.
In graphyne-$N$, benzene rings are connected by 1D polyyne chains (\ce{-C#C-})$_N$ 
to form a triangular lattice.
The system has the point group symmetry $D_{6h}$, which is generated by
inversion $I$, sixfold rotation $C_{6{\bm z}}$ around $\bm{z}$ axis, and twofold rotation $C_{2 {\bm x}}$ around $\bm{x}$. 
Hereafter $C_{n {\bm{q}}}$ denotes an $n$-fold rotation around $\bm{q}$ axis.

We calculate the atomic and electronic structures of graphyne-$N\!=\!1$, 2, 3, and 4 using the first-principles calculation package 
{\sc quantum espresso}~\cite{giannozzi2009} (see Appendix~\ref{sec:cmpt} for the details of calculation).
Figure \ref{fig:band2d} plots the electronic bands in the optimized atomic structure 
along the high symmetry lines in the Brillouin zone illustrated in Fig.~\ref{fig:lattice}(d).
We see that the graphynes from $N=1$ to 4 are all semiconductors with band gaps of a few 100 meV~\cite{narita1998},
while we observe a characteristic even-odd effect in the low-energy band structure.
For even $N$ [Figs.~\ref{fig:band2d}(b) and (d)], a gap minimum occurs at the $\Gamma$ point.
The band edge states are characterized as $E_{2u}$ and $E_{1g}$ representation of point group $D_{6h}$,
which have double degeneracy with odd and even parities, respectively, for inversion $I$.
For odd $N$, on the other hand, the gap minimum is located at $M$ point, where the symmetry is $D_{2h}$ [Figs.~\ref{fig:band2d} (a) and (c)]. 
The band edge states are $A_{u}$ and $B_{3g}$ representation, and they have the same parity under rotation $C_{2\bm{x}}$ while
opposite parities under inversion $I$ and $C_{2\bm{z}}$.

These band structures are well reproduced by single-orbital tight-binding models as shown below.
As graphyne-$N$ is a planar sheet, we have the mirror reflection symmetry $M_z$ with respect to the $xy$-plane,
and hence $p_z$ atomic orbitals with odd mirror parity are decoupled from $s$, $p_x$, and $p_y$ orbitals with even mirror parity.  
The latter three orbitals form the $sp^2$ and  $sp$ covalent bonds in the benzene rings and the 1D chain, respectively,
resulting in a large band gap at the Fermi energy.
Therefore, the band structure around Fermi energy is originating from $p_z$ atomic orbitals.

Based on this argument, we construct a tight-binding model for the $p_z$ atomic orbital on the carbon sites in Fig.~\ref{fig:lattice}.
We take into account only the nearest neighbor hoppings $t_0, t_1, t_2,\cdots$ in Fig.~\ref{fig:lattice},
and define them by
\begin{eqnarray}~\label{eq:hop2d}
	t_i =  t_{\pi} \exp(-a_i/\lambda),
\end{eqnarray}
where $a_i$ is the corresponding carbon-carbon distance obtained from the optimized atomic-structure in the first-principles calculation,
$\lambda=0.9$ \AA\, is the decay length of the hopping integral, and $t_\pi = 13.1$ eV.
The parameters $t_\pi $ and $\lambda$ are determined to reproduce the graphene's energy band \cite{slater1954,nakanishi2001, moon2012, moon2013}.
In Table \ref{table:2D}, we provide the list of the bond distance $a_i$ for graphyne-$N$ with $N=1,2,3,4$. 
The $a_i$ varies from 1.22 \AA \, to 1.42\AA \, depending on the position.
Accordingly, the corresponding hopping integral $t_i$ varies from 2.7 eV to 3.4 eV.

\begin{table}[tb]
	\caption{Numerically obtained atomic distance in the monolayer graphyne-$N$.
	$a_0=x_1-L/2$ and $a_{i}=x_{i+1}-x_i$ ($i>1$) are the distances of carbons in benzene rings and in polyyne chain [see also Fig.~\ref{fig:chain}(a) for the definition].} \label{table:2D}
	\renewcommand{\arraystretch}{1.5}
	{\tabcolsep = 3.3mm
	\begin{tabular*}{85mm}{c|cccccc}
		\hline 
		\hline 
		&   \multicolumn{6}{c}{Atomic distance [\AA]} \\
		$N$ & $a_0$ & $a_1$ & $a_2$ & $a_3$ & $a_{4}$ & $a_5$ \\ \hline
		 1  & 1.41  & 1.39  & 1.22  &       &         &       \\
		 2  & 1.42  & 1.38  & 1.23  & 1.33  &         &       \\
		 3  & 1.41  & 1.39  & 1.22  & 1.33  &  1.23   &       \\
		 4  & 1.41  & 1.38  & 1.23  & 1.32  &  1.24   &  1.31 \\
		\hline 
		\hline
	\end{tabular*}
	}
	\renewcommand{\arraystretch}{1}
\end{table}

In Fig.~\ref{fig:tb2d}, we present the band structures obtained from the nearest-neighbor tight-binding model with Eq.~(\ref{eq:hop2d}).
We see that the tight-binding model captures the main features of DFT band calculations in Fig.~\ref{fig:band2d} around the Fermi energy.

\subsection{Origin of even-odd effect}\label{sec:evenodd}

\begin{figure}[t]
\begin{center}
	\includegraphics[width=85mm]{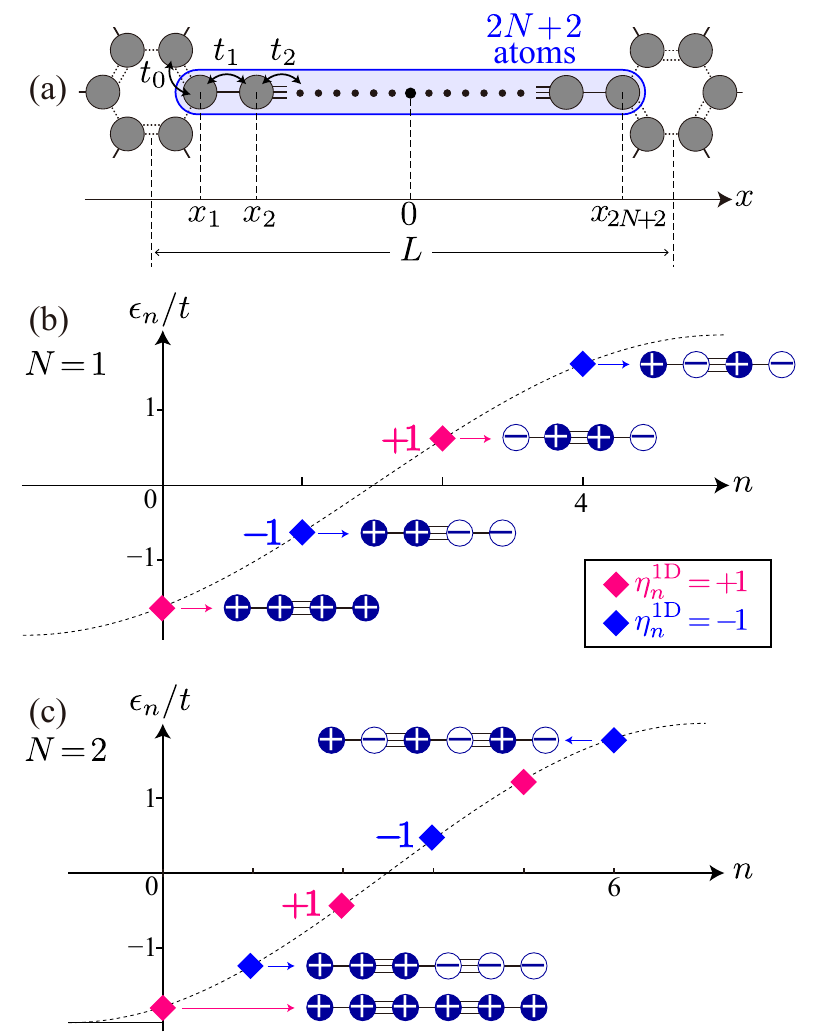}
	\caption{
	(a) 1D atomic chain as a component of graphyne-$N$. (b) The energy levels and the schematic plot of the eigenstates
	 (sign of the wave function on the atomic sites) for chain of $N=1$ and (c) $N=2$.
	Color of each energy level indicates the parity $\eta^{\rm 1D}_n$.
	}\label{fig:chain}
\end{center}
\end{figure}

The even-odd effect in the band structure of graphyne-$N$ can be understood
using the tight-binding model introduced above. 
As depicted in Fig.~\ref{fig:chain}(a),
graphyne-$N$ can be divided into isolated 1D chains consisting of $2N+2$ carbon atoms,
by turning off the hopping $t_0$ in the benzene ring.
The whole system can be viewed as an effective tight-binding model,
with each 1D chain as an effective site, and $t_0$ as hopping between neighboring effective sites.
For simplicity, we assume that the all the distances between neighboring atoms $a_0, a_1, a_2 \cdots$ are equal (denoted by $a$),
and hence all the hopping parameters $t_0, t_1, t_2,\cdots$ are all equal (denoted by $t$).
Here we take the origin at the center of the chain, and define the position of $j$-th atom as $x_j= [j-(2N+3)/2]a$ for $j=1,2,\cdots, 2N+2$.

The Schr\"odinger equation for the isolated chain is given by 
$\epsilon_n \psi_n (x_j) = - t[\psi_n (x_{j+1} )+ \psi_n (x_{j-1})]$. 
The eigen energy of the chain is
\begin{eqnarray}\label{eq:echain}
	\epsilon_{n} = -2t \cos (k_n a),
\end{eqnarray}
with the quantized Bloch wave number
\begin{eqnarray}
	k_n = \frac{ n\!+\!1}{L}\pi,
\end{eqnarray}
where  $n=0,1,\cdots, 2N+1$ and $L=(2N+3)a$, and $a$ is the lattice spacing in the chain.
The level structures for $N=1$ and $2$ are shown in Figs.~\ref{fig:chain}(b) and (c), respectively.
The wave function of the $n$-th eigen state is 
\begin{eqnarray}\label{eq:psichain}
	\psi_{n}(x_j) = \sqrt{\frac{2}{2N+3}} \cos \left(k_n x_j + \frac{n\pi}{2}\right).
\end{eqnarray}
The parity of the 1D wave function, defined by $\psi_n(x_i)=\eta^{\rm 1D}_n \psi_n(-x_i)$,
is crucial to consider the band structure of the graphyne-$N$.
From the Eq.~(\ref{eq:psichain}), the parity of the $n$-th eigenstates is given by
\begin{eqnarray}\label{eq:c2parity}
\eta^{\rm 1D}_n=
\left\{\begin{array}{l}
-1 \hbox{ for odd } n, \\
+1 \hbox{ for even } n.
\end{array}\right. 
\end{eqnarray} 
For odd $N$, 
the states just above and below $\epsilon=0$ has parity $\eta^{\rm 1D}_n=+1$ and $-1$, respectively  [Fig.~\ref{fig:chain}(b)],
while those for even $N$ become  $\eta^{\rm 1D}_n=-1$ and $+1$, respectively [Fig.~\ref{fig:chain}(c)].
\begin{figure}[b]
\begin{center}
	\includegraphics[width=85mm]{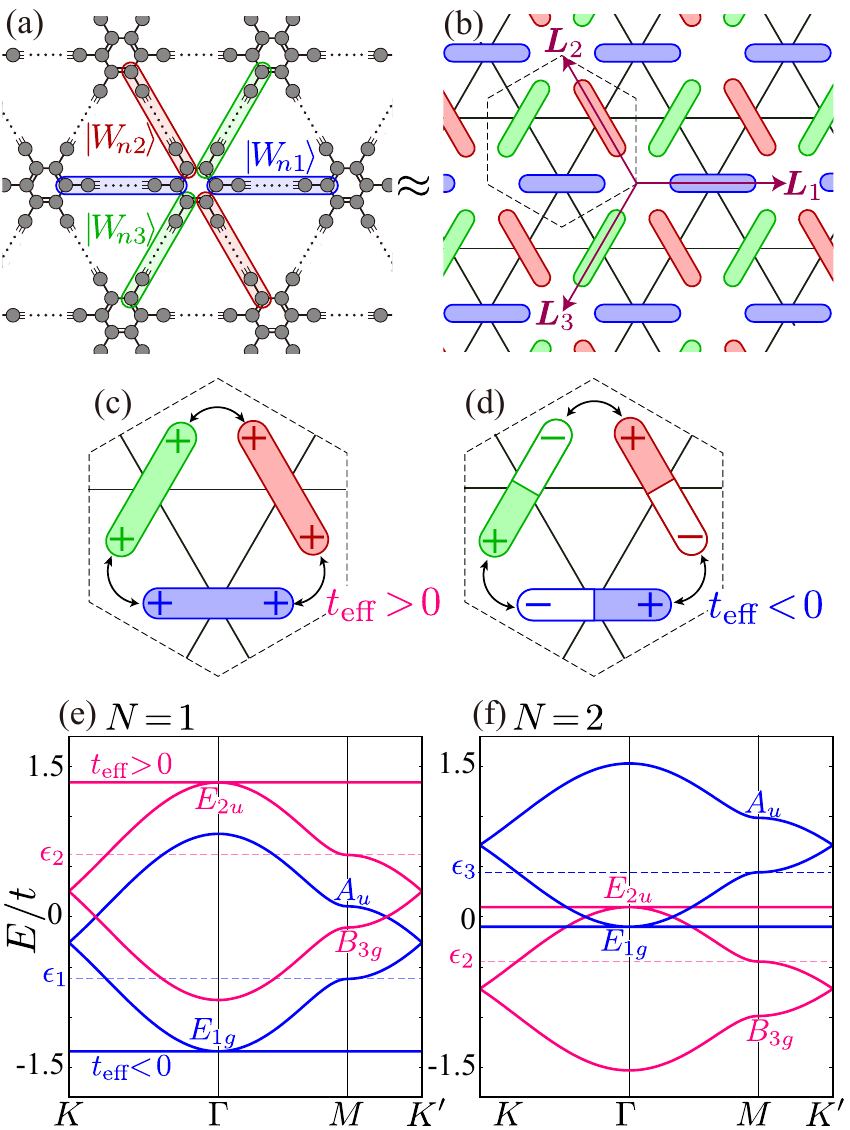}
	\caption{
	(a) Decomposition of graphyne-$N$ into 1D-chain effective sites
	and (b) the corresponding kagome lattice. 
	(c) Schematics of the hopping between neighboring effective orbitals with even parity and (d) odd parity.
	(e) Configuration of the effective kagome bands of $n=N$ and $n=N+1$  for graphyne $N=1$ and (f) $N=2$,
	where the mixing between different $n$'s is neglected.
	}\label{fig:kagome}
\end{center}
\end{figure}

We regard these 1D chains as effective sites. 
They form a kagome lattice in graphyne-$N$ as shown in Figs.~\ref{fig:kagome}(a) and (b). 
We define the primitive lattice vectors $\bm{L}_1$, $\bm{L}_2$, $\bm{L}_3$ as in Fig.~\ref{fig:kagome}(b),
where $\bm{L}_3 = -\bm{L}_1 - \bm{L}_2$.
The unit cell of the kagome lattice includes the three chains along different directions.
These orbitals are expressed as $\left|W_{n,\alpha}(\bm{R}_{\alpha}) \right>$ [Figs.~\ref{fig:kagome}(a)],
where $n$ corresponds to the $n$-th eigenstate the 1D chain,
$\alpha = 1,2,3$ is the index for the direction of chain,
and $\bm{R}_{\alpha} = m_1 \bm{L}_{1} +m_2 \bm{L}_{2} +  \bm{L}_{\alpha}/2$ ($m_1, m_2$: integers)
is the central position of the orbital $\alpha$.
Specifically, $\left|W_{n,\alpha}(\bm{R}_{\alpha}) \right>$ is given by arranging the wave amplitudes of $\psi_n(x_1), \psi_n(x_2), \psi_n(x_3), \cdots$ 
on the atomic sites from one end to the other in the direction of $\bm{L}_\alpha$.

The even-odd feature of the band structure can be captured
by an approximate model which takes only the matrix elements between the same $n$, while neglecting those between different $n$'s.
In this approximation, the Bloch Hamiltonian of the sector $n$ is given by $H_{n,\alpha,\beta}(\bm{k})=\langle w_{n,\alpha}(\bm{k}) |H| w_{n,\beta}(\bm{k}) \rangle$, 
where the Bloch state of $n$-th orbital is defined as 
\begin{eqnarray}\label{eq:bloch}
 |w_{n, \alpha} ({\bm k}) \rangle = \sum_{\bm{R} \in \bm{R}_{\alpha}} \exp(i\bm{k}\cdot \bm{R}) | W_{n,\alpha}(\bm{R})\rangle.
\end{eqnarray}
When turning on the hopping  in the benzene ring, the ends of the neighboring chains are coupled with each other.
It gives the nearest neighbor hopping in the effective kagome lattice.
The Bloch Hamiltonian is expressed in a matrix form with the indexes $\alpha, \beta$ as 
\begin{eqnarray}\label{eq:hkagome}
	H_{n}({\bm{k}}) = \epsilon_n \hat{1} 
	-t_{\mathrm{eff}}
	\left(\begin{array}{ccc} 
		0 & \cos \theta_{3}(\bm k) & \cos \theta_{2}(\bm k) \\
		\cos \theta_{3}(\bm k) & 0 & \cos \theta_{1}(\bm k) \\
		\cos \theta_{2}(\bm k) & \cos \theta_{1}(\bm k) &  0
	\end{array}\right),
\end{eqnarray}
where $\theta_{\alpha}(\bm{k})=\bm{k}\cdot\bm{L}_\alpha/2$ and $\hat{1}$ is a $3\times3$ unit matrix,
and $t_{\rm eff}$ is the nearest neighbor hopping in the kagome lattice, which is given by 
\begin{eqnarray}
t_{\mathrm{eff}}(n) &=& t \,\psi_n(-x_1)\psi_n(x_1) \nn\\
&=& \eta^{\rm 1D}_{n} \, \frac{2t}{2N+3}\cos^2\frac{(2N-2n+1)\pi}{2(2N+3)}. 
\end{eqnarray}
An important point here is that the parity of the 1D-chain wave function determines the sign of $t_{\mathrm{eff}}(n)$.
As can be seen in Fig.~\ref{fig:kagome}(c), when the orbital has even parity $\eta^{\rm 1D}_{n}=+1$,  
the effective orbitals of the two neighboring chains have the same sign at the contact point.
Therefore, the effective hopping between orbitals is $t_{\mathrm{eff}}(n) > 0$ (noting that $t > 0$).
In the same manner, we obtain $t_{\mathrm{eff}}(n)< 0$ for the orbital with odd parity $\eta^{\rm 1D}_{n}=-1$ [Fig.~\ref{fig:kagome}(d)].
Note that the one dimensional modes and the associated kagome lattice was also proposed in polymerized triptycene~\cite{mizoguchi2019}. 

The eigenvalues of Eq.~(\ref{eq:hkagome}) is 
\begin{eqnarray}\label{eq:ekagome}
	E_{n \bm{k}} &=& \epsilon_n  + 2t_{\mathrm{eff}}, \nn\\ 
	&&\epsilon_n  -  t_{\mathrm{eff}}\Big\{\!1\!\pm\! \Big[\sum_\alpha [1\!+\!2\cos 2\theta_{\alpha}(\bm{k})]\Big]^{\frac{1}{2}}\!\Big\},
\end{eqnarray}
where the former represents a flat band, and the latter represents a pair of dispersive bands analog to graphene.

The effective kagome lattice clearly explains the even-odd effect of graphyne-$N$.
For odd $N$, the 1D-chain eigenstates of  just above and below the zero energy 
have even and odd parity, respectively, as shown in Fig.~\ref{fig:chain}(b). 
These two orbitals give rise to a pair of kagome band near the Fermi energy as depicted in Fig.~\ref{fig:kagome}(e).
There the blue curve originates from $n=N$ state with the odd parity,
and hence it has a flat band in the lower energy side,
while the magenta curve is from $n=N+1$ state with the even parity 
and hence the flat band appear in the higher energy side.
In this system, the energy difference between 1D-chain
orbitals, $\epsilon_{N+1}-\epsilon_{N}=2|\epsilon_{N}|$, is smaller than the kagome band width $8|t_{\mathrm{eff}}(N)|$ for any $N\ge1$,
 so that the two kagome clusters overlap with each other, giving band crossings between $K$ and $\Gamma$, and also between $M$ and $K'$.

In the real band structure, a finite mixing between different $n$'s neglected in Eq.\ (\ref{eq:hkagome})
induces a mini gap at these crossing points, resulting in semiconducting band structures as shown in Figs.\ \ref{fig:band2d}(a) and \ref{fig:band2d}(c).
The symmetry character at the high symmetric points also depends on the parity of orbitals.
For the even (odd) parity band in Fig.~\ref{fig:kagome}(e), 
the twofold degenerate states at $\Gamma$ point belong to $E_{2u}$ ($E_{1g}$) representations and 
the saddle point state at $M$ with $E_{n}(M)=\epsilon_n-2t_{\mathrm{eff}}$ to $B_{3g}$ and ($A_{u}$) 
[see Appendix~\ref{sec:kagome} for details].

In even-$N$, in contrast, the sign of $t_{\mathrm{eff}}$ becomes opposite, 
and therefore the low energy kagome bands are arranged such that 
flat-band sides are faced to each other as in Fig.~\ref{fig:kagome}(f).
The two sets of bands overlap in such a way that 
the $E_{2u}$ states from the lower band cluster (magenta bands) has higher energy than $E_{2g}$ from the upper band cluster (blue).
The mixing between different $n$'s  opens a band gap at the crossing points as in Figs.\ \ref{fig:band2d}(b) and \ref{fig:band2d}(d).
As a result, there is a double band inversion of twofold degenerate state $E_{1g}$ and $E_{2u}$, and hence the system is in a topological insulating phase.
In fact, it has been pointed out that the monolayer graphyne-2 is the second order topological insulator associated with this double band inversion~\cite{lee2020}.
Note that the similar topological phase also appears Kekul\'e distorted honeycomb lattice~\cite{wu2016,kariyado2017} and 
its counterpart in the photonic crystals~\cite{wu2015}.

\begin{figure}[b]
\begin{center}
	\includegraphics[width=85mm]{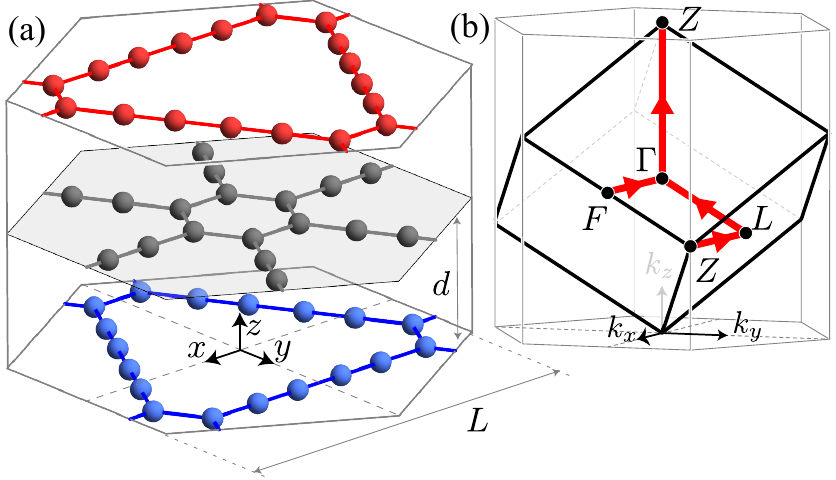}
	\caption{
	(a) Lattice structure of ABC-stacked graphyne-2 and (b) the corresponding Brillouin zone.
	}\label{fig:abclat2}
\end{center}
\end{figure}

\begin{figure}[t]
\begin{center}
	\includegraphics[width=85mm]{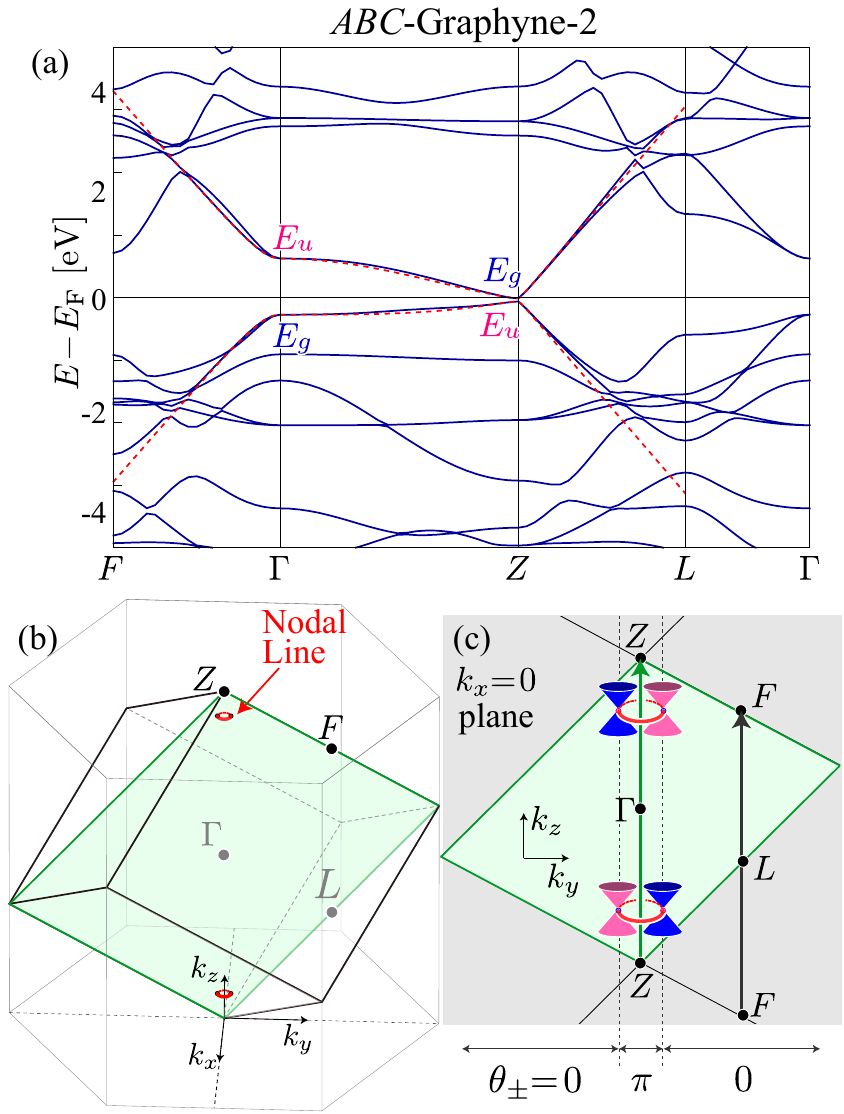}
	\caption{ 
	(a) Band structure of ABC stacked graphyne-2 obtained from the DFT calculation.
	The dashed line is the dispersion relation calculated by  the effective model Eq.~(\ref{eq:h4x4}). 
	(b) The nodal line structure in the three-dimensional momentum space and (c) on $k_x=0$ plane. 
	In (c), the blue and magenta Dirac cones correspond to the mirror eigenvalues $\eta(M_{x})=\pm1$, respectively.
	}\label{fig:abcband2}
\end{center}
\end{figure}

\section{ABC-stacked graphynes (Even $N$)}\label{sec:abceven}
The observation of the even-odd effects in monolayer graphyne-$N$ 
motivates us to systematically study the electronic structures of three dimensional graphyne-$N$.
As in graphene, there are various types of stacking configurations in graphyne multilayers.
Recently, X-ray diffraction experiment report that 
graphyne of $N=2$ obtained by alkyne-alkyne homocoupling reaction takes ABC (rhombohedral) stacking structure~\cite{matsuoka2017}. 
We investigate the electronic and topological properties of ABC-stacked graphyne of $N=2,4$ in this section,
and that of $N=1,3$ in the next section.

\subsection{ABC stacked graphyne-2}~\label{sec:abc2}
We consider the band structure and the topological property of ABC-stacked graphyne-2.
In the previous works, it was found that the system has a nodal line near the Fermi energy~\cite{nomura2018},
and it is characterized by a nontrivial $\mathbb{Z}_{2}$ monopole charge associated with the double band inversion \cite{ahn2019}.
In the following, we argue the origin of the double band inversion and the emergent nodal line from the 
viewpoint of the full crystalline symmetry.

\begin{figure}[b]
\begin{center}
	\includegraphics[width=85mm]{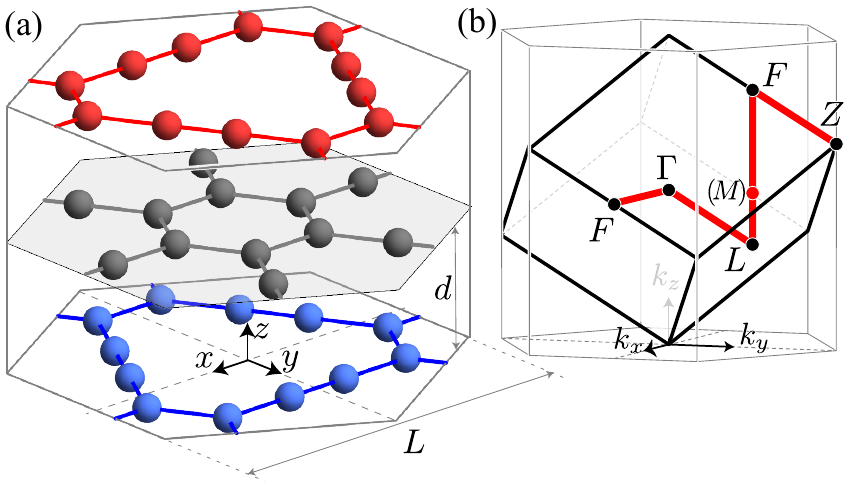}
	\caption{
	(a) Lattice structure of ABC-stacked graphyne-1 and (b) the corresponding Brillouin zone.
	}\label{fig:abclat1}
\end{center}
\end{figure}

The lattice structure of ABC-graphyne-2 is shown in Fig.~\ref{fig:abclat2}(a). 
The primitive lattice vectors are given by 
$\bm{L}_1=\frac{L}{2\sqrt{3}} \hat{\bm y}+d \hat{\bm{z}}$, 
$\bm{L}_2=-\frac{L}{2} \hat{\bm x}-\frac{L}{2\sqrt{3}} \hat{\bm y}+d \hat{\bm{z}}$, and
$\bm{L}_3= \frac{L}{2} \hat{\bm x}-\frac{L}{2\sqrt{3}} \hat{\bm y}+d \hat{\bm{z}}$.
The first Brillouin zone is given by a rhombohedron [Fig.~\ref{fig:abclat1}(b)].
The point group symmetry is $D_{3d}$.
Compared to the monolayer counterpart, the sixfold rotational symmetry $C_{6\bm{z}}$ is reduced to threefold symmetry $C_{3\bm{z}}$,
and also $C_{2\bm{y}}$ is broken. 

Figure~\ref{fig:abcband2}(a) displays the band structure of ABC graphyne-2, which is obtained from the DFT calculation
with the optimized lattice structure in Fig.~\ref{fig:abclat2}(a). 
The nodal lines are not located on the high-symmetry path in Fig.~\ref{fig:abcband2}(a),
but they are at off-center momenta surrounding $k_z$-axis, as depicted in Fig.~\ref{fig:abcband2}(b)~\cite{nomura2018,ahn2019}.
The key factors for the emergence of the nodal lines are the band degeneracy and the double band inversion between the $\Gamma$ and $Z$ points. 
These two points have the $D_{3d}$ group symmetry,
and the two-fold degenerate states near the Fermi energy 
are characterized by  $E_{g}$ and $E_{u}$  representations, 
which are originating from $E_{1g}$ and $E_{2u}$ states in the monolayer system. 
At the $\Gamma$ point, $E_{u}$ is higher than $E_{g}$ in energy,
while at $Z$ point, $E_{g}$ is higher than $E_{u}$.
Since $E_{g}$ and $E_{u}$ are even and odd under the space inversion $I$, respectively,
we have a double band inversion between $\Gamma$ and $Z$ points.

The $k_x=0$ plane including $\Gamma$ and $Z$ point is invariant under mirror reflection $M_x$,
where the bands are labeled by the mirror eigenvalue $\eta(M_x)=\pm 1$.
At $\Gamma$ and $Z$, each of the doublet states $E_{u}$ and $E_{g}$ 
is composed of opposite mirror eigenvalues $\eta(M_x)=+1$ and $-1$.
Therefore, the two sectors of $\eta(M_x)=+1$ and $-1$ 
simultaneously have a band inversion in the eigenvalue of inversion $I$ between $\Gamma$ and $Z$ point.

On the $k_x=0$ plane, we can define the Berry phase (Zak phase) $\theta_{\pm}(k_y)$ for each sector of $\eta(M_x)=\pm 1$,
by integrating the Berry connection along $k_z$ axis for a span of the Brillouin zone with $k_y$ fixed [see Fig.~\ref{fig:abcband2}(c)].
The $\theta_{\pm}(k_y)$ is quantized to $0$ or $\pi$ because of the inversion and time-reversal symmetry of the present system.
On the $Z \Gamma Z$ line $(k_y=0)$, the band inversion in each mirror sector ensures 
$\theta_{\pm}(0)=\pi$~\cite{fu2007}. 
On the line $FLF$ $(k_y=\pi/L)$, in contrast, there are no band inversion because of a large band gap, 
and thus $\theta_{\pm}(\pi/L)=0$.
Between the two lines $Z\Gamma Z$ and $FLF$, therefore, 
there must be a jump of the Berry phase from $\theta_{\pm}(k_y)=\pi$ to $0$, 
at which energy gap closes in the corresponding sector of $\eta(M_x)=\pm$. 
This is the cross section of the nodal line and the $k_x=0$ plane.
The $\eta(M_x)=\pm$ sectors have gap closing generally at different positions in the momentum space,
as illustrated by the magenta and blue Dirac cones, respectively, in Fig.~\ref{fig:abcband2}(c).
The position of the gap closing point will be argued later in more detail.
The Berry phase on an infinitesimal closed path encircling a single gap closing point is $\pi$.
As the Berry phase on a closed path is always quantized to $0$ or $\pi$ due to the inversion and time-reversal symmetry, 
the gap closing point persists even away from $k_x=0$ plane.
This explains the nodal lines in the three-dimensional momentum space.

In the following, we derive an effective $4\times 4$ low-energy Hamiltonian 
from a symmetry-based consideration for the states $E_{g}$ and $E_{u}$ at the $\Gamma$ points.
We define the four-dimensional basis as $|\alpha,\beta\rangle$ 
where $\alpha=\pm$ represent $E_g$ and $E_{u}$, respectively
and $\beta=\pm$ is the degree of freedom in the twofold degeneracy,
corresponding to the sign of the angular momentum [see Appendix~\ref{sec:fourband}].
In this basis, a reducible $4\times 4$ representation of the twofold rotation and improper six fold rotation generating of $D_{3d}$ are
\begin{eqnarray}
	 C_{2 \bm{x}} = \sigma_x \tau_0, \quad \hbox{and} \quad S_{6} = e^{2\pi i/3 \sigma_z} \tau_z.
\end{eqnarray}
Here, $\tau_{i}$ with $i=0, x,y,z$ are the unit matrix and the Pauli matrices acting 
on $\alpha$, while $\sigma_{i}$ is on $\beta$.
The time reversal operator is expressed as
\begin{eqnarray}
	&\mathcal{T} =  \sigma_x \tau_z \mathcal{K},
\end{eqnarray}
where $\mathcal{K}$ is the complex conjugate operator.

\begin{table*}[t]
	\caption{Model parameters in Eq.~(\ref{eq:fourier}) for graphyne-2 and 4 which reproduce the band structure obtained from DFT. 
	The other parameters are set to zero. $N=4'$ stands for the graphyne-4 under the pressure $P\approx3.8$ GPa.} \label{table:abceven}
	\renewcommand{\arraystretch}{1.5}
	{\tabcolsep = 5.5mm
	\begin{tabular*}{180mm}{lrrrrrrrrr}
		\hline 
		\hline 
		$N$& $\tilde{m}_{00}$ & $\tilde{m}_{01}$ & $\tilde{m}_{02}$ &$\tilde{m}_{10}$ & $\tilde{m}_{11}$ & $\tilde{m}_{12}$ & $\tilde{m}'_{1}$ & $\tilde{m}'_{2}$ & $\tilde{v}_{0}/L$  \\
		\hline
		2  & 0 & $0.125$ & $0$ & 0.213 & $0.233$ & $0$ & $-0.23$ & $0$   & $0.491$     \\
		4  & 0.07 & $-0.02$ & $-0.02$ & 0.24 & $-0.16$ & $ 0.01$ & $0.17$ & 0.17 & $0.3$  \\
		$4'$ & 0.16 & $-0.06$ & $-0.1$  & 0.13 & $-0.36$ & $ 0.1$ & $0.41$ &-0.17 & $0.3$  \\
		\hline 
		\hline
	\end{tabular*}
	}
	\renewcommand{\arraystretch}{1}
\end{table*}

The symmetry requirements for the effective Hamiltonian $H(\bm{k})$ are then written as
$S_6H(\bm{k})S_6^{-1}=H(D_{S_{6}}[\bm{k}])$, 
$C_{2\bm{x}}H(\bm{k})C_{2\bm{x}}^{-1}=H(D_{C_{2\bm{x}}}[\bm{k}])$, 
and $\mathcal{T}H(\bm{k})\mathcal{T}^{-1}=H(-\bm{k})$.
Up to the first order of $k_x$ and $k_y$, $H(\bm{k})$ is uniquely determined as
\begin{eqnarray}\label{eq:h4x4}
	H(\bm{k}) &=& m_0 \!+\! m_1\tau_z   \!+\! v \tau_x (\sigma_x  k_x  \!+\! \sigma_y k_y)  \nn\\
	 &&  + m' \tau_y\sigma_z \!+\! (v'_1 \!+\! v'_2 \tau_z ) (\sigma_y k_x \!-\! \sigma_x k_y),
\end{eqnarray}
where $m_0(k_z)$, $m_1(k_z)$ and $v(k_z)$ are even function of $k_z$ 
and $m'(k_z)$, $v'_1(k_z)$, and $v'_2(k_z)$ are odd function (see Appendix~\ref{sec:fourband} for derivation).

In the Fourier series of $k_z$, we can write 
\begin{eqnarray}\label{eq:fourier}
 m_i(k_z) &=& \sum_{q=0} \tilde{m}_{i q} \cos(q k_z d), \nn\\ 
 v(k_z) & = &  \sum_{q=0} \tilde{v}_{q} \cos( q k_z d), \nn\\ 
 m'(k_z) & = &  \sum_{q=1} \tilde{m}'_q \sin( q k_z d), \nn\\  
 v'_i(k_z) &=&  \sum_{q=1} \tilde{v}'_{i q}  \sin( qk_z d).
\end{eqnarray}
The model within the first harmonics (i.e., only the terms with $q \leq 1$)
well reproduces the band structure of first-principles calculation [see Fig.~\ref{fig:abcband2}(a)],
by choosing the parameters listed in Table~\ref{table:abceven}.

The model without $v'_i(k_z)$ terms in Eq.~(\ref{eq:h4x4}) is equivalent to the effective Hamiltonian derived in Ref.~\cite{nomura2018}. 
Actually $v'_i(k_z)$ is irrelevant in the graphyne-2, because the low-energy bands are located near $Z$ point, where $\sin(k_z d)$ is small.
The effective model Eq.~(\ref{eq:h4x4}) also applies to the graphynes with even $N$'s, which have the same crystal symmetry. 
Also, the model with $k_z$-dependent terms neglected
(i.e., $\tilde{m}_{i q}=\tilde{v}_{q}=\tilde{m}_{q}'=\tilde{v}'_{q}=0$ for $q\ge 1$) describes the monolayer graphynes of even $N$.
The fitted parameters for monolayer graphene $N=2$,and $4$ are given in the Table~\ref{table:monoeven}
which give dashed curves in Figs.~\ref{fig:band2d}(b) and (d).

The model Eq.~(\ref{eq:h4x4}) is analytically solvable. 
Let us first consider $k_x=0$ plane which is invariant under the mirror reflection $M_x$.
The mirror reflection operator acting on the effective Hamiltonian Eq.~(\ref{eq:h4x4}) is given by
\begin{eqnarray}
	M_x= S_6^3 C_{2\bm{x}}=\sigma_x\tau_z. 
\end{eqnarray}
It is diagonalized to $M'_{x}=UM_{x}U^\dag= \tau_z$ by $U=\frac{1}{4}(-\tau_+\sigma_x+\tau_-)(\tau_x\sigma_++\sigma_-)e^{-i(\pi/4)\sigma_y}$, 
with $\sigma_\pm=\sigma_x\pm i\sigma_y$ and $\tau_\pm=\tau_x\pm i\tau_y$.
By using this unitary transformation, we can block diagonalize the effective Hamiltonian Eq.~(\ref{eq:h4x4}) as
\begin{eqnarray}
	UH(0,k_y,k_z)U^\dag = H_{+} \oplus H_{-}
\end{eqnarray}
with 
\begin{eqnarray}
	H_{\eta}\!=\! m_0 \!-\! \eta v'_2 k_y 
	\!-\! (m_1 \!-\! \eta v'_1 k_y) \sigma_z  
	\!-\! (m'\!-\! \eta v k_y) \sigma_y. 
\end{eqnarray}
where $\eta$ indicates the eigenvalue of mirror operator $\eta(M_x)=\pm 1$. 
By diagonalizing $H_{\eta}$, the eigenenergy in $k_x=0$ is obtained as
\begin{eqnarray}\label{eq:Em}
	E_{\pm,\eta}  \!=\! m_0 \!-\! \eta v'_2 k_y 
	\!\pm\! \sqrt{(m_1 \!-\! \eta v'_1 k_y)^2 \!+\! (m'\!-\! \eta v k_y)^2}
\end{eqnarray}

\begin{table}[b]
	\caption{Model parameters in Eq.~(\ref{eq:fourier}) for graphyne-2 and 4 to reproduce the band structure obtained from the DFT calculation.} \label{table:monoeven}
	\renewcommand{\arraystretch}{1.5}
	{\tabcolsep = 8.0mm
	\begin{tabular*}{85mm}{cccc}
		\hline 
		\hline 
		$N$& $\tilde{m}_{00}$ & $\tilde{m}_{10}$ & $\tilde{v}_0/L$ \\
		\hline 
		2& $0.15$ & 0.2  & 0.5 \\
		4& $0.2$ & 0.25  & 0.3 \\
		\hline 
		\hline
	\end{tabular*}
	}
	\renewcommand{\arraystretch}{1}
\end{table}

From Eq.~(\ref{eq:Em}) we obtain the intersection of nodal line and $k_x=0$ plane.
For each sector of $\eta(M_x)=\pm 1$, the gap closing point on the $k_x=0$ plane is obtained by
$E_{+,\eta} = E_{-,\eta}$, or
\begin{eqnarray}\label{eq:nleq}
	\left\{\begin{array}{l}
	m_1(k_z) - \eta v'_1(k_z) k_y = 0 \\
	m'(k_z) - \eta v(k_z) k_y = 0. 
	\end{array}\right.
\end{eqnarray}
We can show that solutions of the Eq.~(\ref{eq:nleq}) exist
when the sign of $m_1(k_z)$ changes between $k_z d=0$ and $\pi$,
namely a band inversion of the $E_{g}$ and $E_{u}$ state takes place between $\Gamma$ and $Z$ points.
It is clear that if there are solutions of Eq.~(\ref{eq:nleq}) $\bm{k}=\pm(0,k_{y0},k_{z0})$ for $\eta(M_x)=+1$, 
we always have other solutions $\bm{k}=\pm(0, -k_{y0},k_{z0})$ for $\eta(M_x)=-1$.
In Fig.~\ref{fig:abcband2}(c), we schematically illustrate such gap closing with $\eta(M_x)=+1$ and $-1$ as magenta and blue cones, respectively.
In addition, the model Eq.~(\ref{eq:h4x4}) has an effective rotation symmetry,
\begin{eqnarray}
	H(D_{\theta}[\bm{k}]) = e^{-i\theta \sigma_z/2} H(\bm{k}) e^{i\theta \sigma_z/2},
\end{eqnarray}
where $D_\theta[\bm{k}]$ represents 
the rotation of $\bm{k}$ along $z$-axis by angle $\theta$.
Therefore, the trajectory of the gap-closing point 
forms a circle parallel to $k_xk_y$ plane 
as shown in Figs.~\ref{fig:abcband2}(b) and (c).

\begin{figure}[b]
\begin{center}
	\includegraphics[width=85mm]{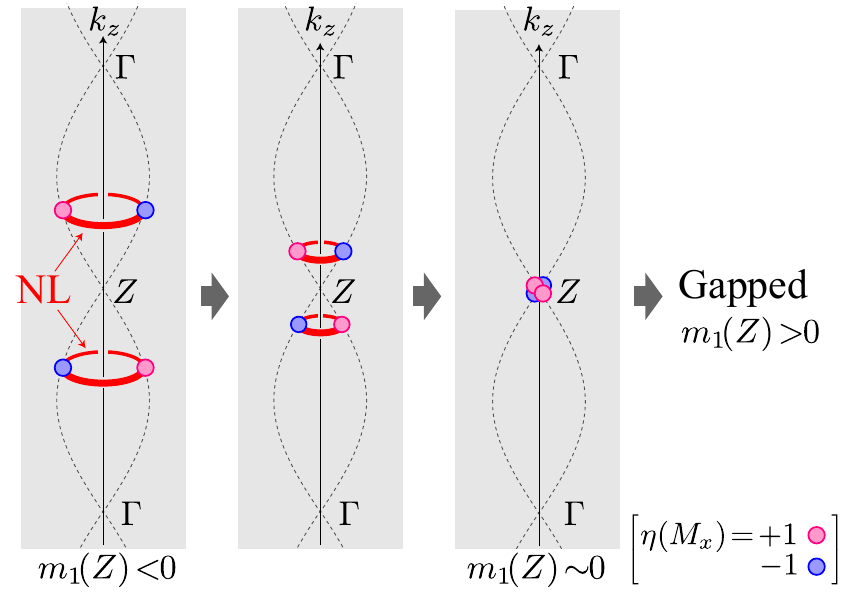}
	\caption{ 
	Pair annihilation of the nodal lines in graphyne-2.
	Blue and magenta dots represent the Dirac points on $k_x=0$ plane with $\eta(M_{x})=+1$ and $-1$, respectively, 
	which are the intersecting points of nodal ring and $k_x=0$ plane.
	Dashed curves are the trajectory of the Dirac points towards the pair annihilation at $Z$ point,
	which is obtained as solutions of the second line of Eq.~(\ref{eq:nleq}) with $m'(k_z)=u' \sin (k_z d)$ and $v(k_x)  = v_0$.
	}\label{fig:gapout}
\end{center}
\end{figure}

Each single nodal ring cannot disappear in its own even when shrunk to a point.
This stability of the nodal line in ABC-graphyne-2 is ensured by a nonzero $Z_2$ monopole charge defined 
by the time-reversal and inversion symmetry~\cite{ahn2019}.
Meanwhile, we can also understand this stability by using crystalline symmetry.
We see in Fig.~\ref{fig:abcband2}(c) that intersection of the horizontal nodal ring and $k_x=0$ plane forms
two Dirac cones in $k_y$-$k_z$ plane. 
As these two cones belong to the sectors with opposite mirror eigenvalue $\eta(M_x)$, 
the mass term is not allowed even when they overlap with each other.
To gap out this nodal ring, one needs to merge the two Dirac cones with the same mirror eigenvalues $\eta(M_x)$, as shown in Fig.~\ref{fig:gapout}. 
With changing the model parameter from $m_1(Z)<0$ to $m_1(Z)>0$ (from $m_1(\Gamma)>0$ to $m_1(\Gamma)<0$) continuously, 
the Dirac points associated with two nodal rings move along dashed curve in Fig.~\ref{fig:gapout} 
and annihilate in pair at $Z$ ($\Gamma$) points when the band inversion of $E_{g}$ and $E_{u}$ is removed.

\begin{figure}[b]
\begin{center}
	\includegraphics[width=85mm]{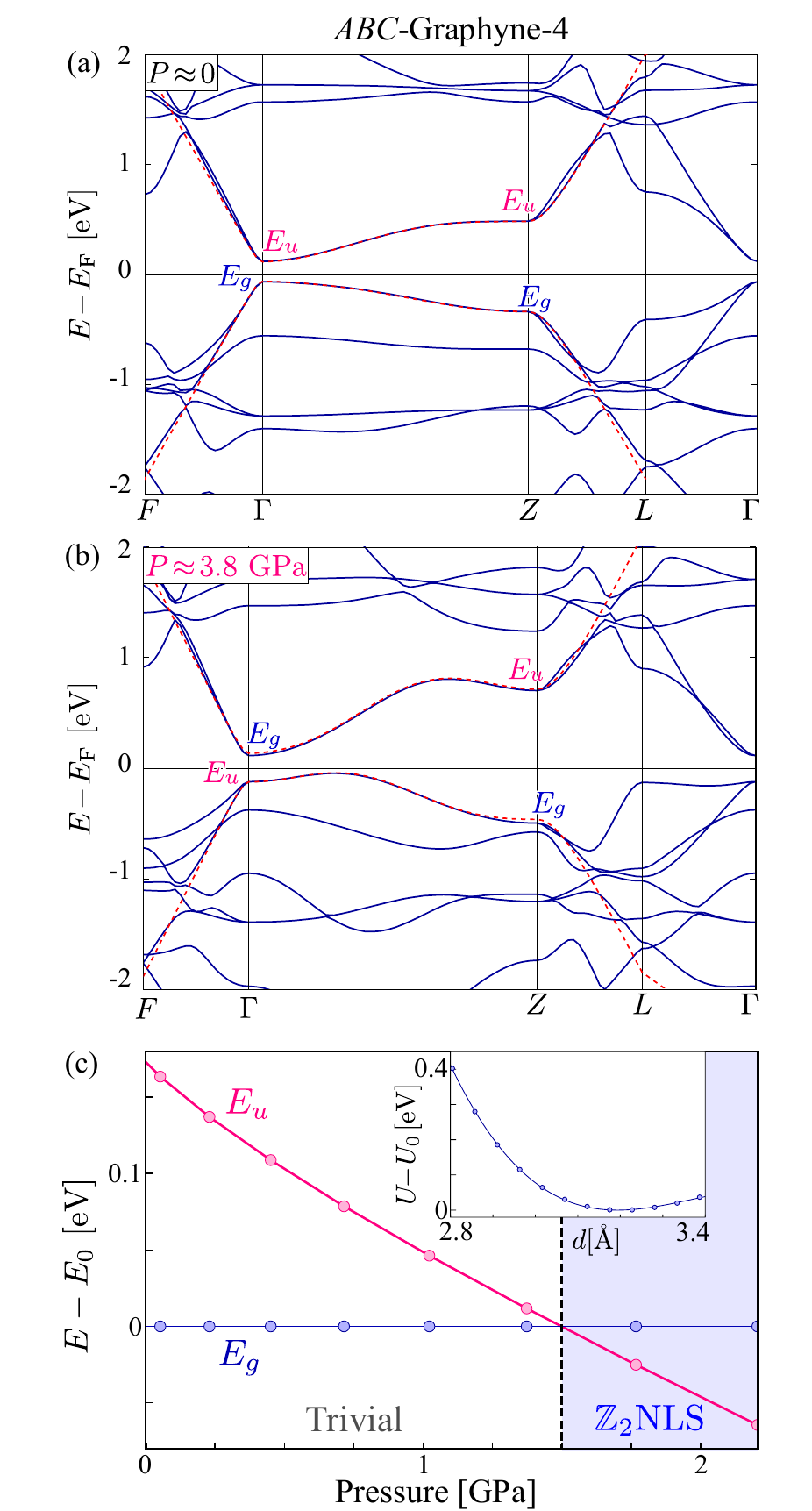}
	\caption{ 
	Energy spectrum of ABC-stacked graphyne-4 obtained from DFT calculation,
	with interlayer distance (a) $d \approx 3.25$ \AA\, ($P\approx 0$) and (b) 2.66 \AA\, ($P\approx 3.8$ GPa).
	(c) Energy difference between $E_g$ and $E_u$ states at $\Gamma$ point as a function of pressure. 
	Inset shows the  total energy of the system (measured from its minimum $U_0$) as a function of the interlayer distance $d$.
	}\label{fig:abc4}
\end{center}
\end{figure}

\begin{figure}[b]
\begin{center}
	\includegraphics[width=85mm]{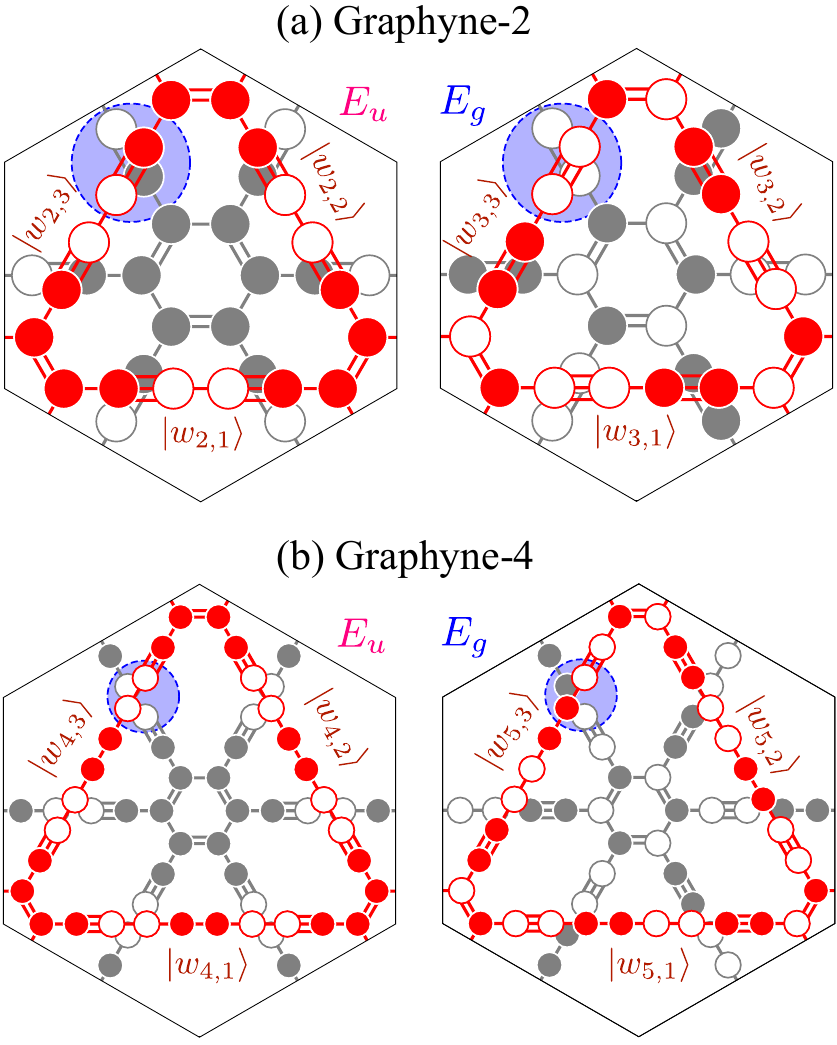}
	\caption{ 
	Atomic structure and wave functions on two neighboring layers of ABC-stacked (a) graphyne-2 and (b) graphyne-4. 
	Red and gray circles represent carbon atoms on layers at $z=+d$ and $z=0$, respectively,
	and open and filled circles indicates positive and negative sign of atomic wave function $\psi_{n}(x_j)$ from Eq.~(\ref{eq:psichain}). 
	Dashed blue circle represents the atomic overlap region with significant interlayer coupling.
	}\label{fig:stack24}
\end{center}
\end{figure}

\subsection{ABC stacked graphyne-4}~\label{sec:abc4}

Figure~\ref{fig:abc4}(a) shows the energy band for ABC-stacked graphyne-4, which is obtained by a similar DFT calculation
with the structural optimization.
Unlike the graphyne-2, the graphyne-4 has no band inversion and hence it is a topologically-trivial semiconductor.
Still, the low-energy band structure can be described by the effective continuum model (\ref{eq:h4x4}),
as it is based on the crystal symmetry of even-$N$ graphyne.
In Fig.~\ref{fig:abc4}(a), the dashed curves are given by the effective model 
upto the second harmonics [$q\le 2$ in Eq.~(\ref{eq:fourier})], with the parameters listed in the Table~\ref{table:abceven}.

In terms of the effective model, the band inversion takes place when $|\tilde{m}_{11}| > |\tilde{m}_{10}+\tilde{m}_{12}|$.
The parameter $\tilde{m}_{11}$ represents the interlayer hopping between the neighboring graphyne layers, 
which occurs in the atomic overlap region illustrated as a dashed circle in Fig.~\ref{fig:stack24}.
We see that the relative size of the dashed circle to the whole unit cell (hexagon)
becomes smaller for larger $N$, resulting in a weaker interlayer coupling in the electronic system.
Due to  the smallness of the interlayer coupling (hence that of $\tilde{m}_{11}$), the graphyne-4 remains a trivial semiconductor.

In van der Waals layered materials, however, the interlayer coupling is highly sensitive to the external pressure,
as it reduces the interlayer distance \cite{proctor2009,nicolle2011, yankowitz2016,ychen2017,yankowitz2018,yankowitz2019}.
Actually, we can show that applying a pressure to ABC graphyne-4 enhances the interlayer coupling, 
and causes a topological phase transition
to a nodal-line semimetal.
To study this effect induced by uniaxial pressure, we carry out the DFT band calculation by changing interlayer distance $d$ systematically.
Under the fixed $d$, the in-plane lattice constant $L$ and the atomic position are determined to minimize thetotal energy, 
and the total energy profile is obtained as a function of $d$ as presented in inset of Fig.~\ref{fig:abc4}(c).
We estimate the corresponding pressure by 
\begin{eqnarray}
	P(d)=-\frac{1}{S} \frac{\partial U}{\partial d},
\end{eqnarray}
where $U$ is the numerically obtained total energy and $S=\sqrt{3}L^2/2$ is size of unit cell projected on $xy$ plane.
Figure~\ref{fig:abc4}(b) shows the band structure of ABC stacked graphyne-4 under $P\approx 3.8$ GPa. 
We see that the pressure induces the band inversion between $E_{g}$ and $E_{u}$ states at the $\Gamma$ point,
resulting in a $\mathbb{Z}_2$ nontrivial nodal-line phase similar to graphyne-2. 
Figure~\ref{fig:abc4}(c) shows the relative energy from $E_g$ state to $E_u$ state at $\Gamma$ point,
where we see that the phase transition takes place at $P\approx 1.5$ GPa. 
The ABC-stacked graphyne-4 is a good platform for the nodal-line semimetal with $\mathbb{Z}_2$ monopole controllable by the pressure.

The effective continuum model Eq.~(\ref{eq:h4x4}) can also describe the band structure under pressure.
The dashed curve in Fig.~\ref{fig:abc4}(b) is actually obtained by the parameters as $N=4'$ in Table~\ref{table:abceven}.
In the model, the condition for the band inversion $|\tilde{m}_{11}| > |\tilde{m}_{10}+\tilde{m}_{12}|$ is satisfied due to the enhancement of interlayer hopping.
We also notice that the pressure distorts the energy bands on $\Gamma Z$ line in Fig.~\ref{fig:abc4},
and this is caused by the enhancement of the second harmonics $\tilde{m}_{02}$ and $\tilde{m}_{12}$, originating from 
the hopping between the next neighboring layers.

As mentioned, the band inversion of graphene-4 takes place at the $\Gamma$ point [Fig.~\ref{fig:abc4}(b)], 
in contrast to graphyne-2 having the inversion at $Z$ point [Fig.~\ref{fig:abcband2}(a)].
In the effective model, the position of the band inversion is attributed to the sign of the parameter $\tilde{m}_{11}$, 
which is $\tilde{m}_{11}<0$ for ABC-graphyne-4 while $\tilde{m}_{11}>0$ for ABC-graphyne-2.
The sign of $\tilde{m}_{11}$ is related with the in-plane electronic structure of $E_g$ and $E_u$ states
consisting of the Bloch wave function $|w_{N,\alpha}(0)\rangle$ and $|w_{N\!+\!1,\alpha}(0)\rangle$ of 1D chain given by~(\ref{eq:bloch})
[see also Fig.~\ref{fig:chain}(c)].
The neighboring interlayer hopping is determined by the relative sign of the wave functions of the two neighboring layers,
in an atomic overlap region [dashed circle of Fig.~\ref{fig:stack24}].
We see that the relative signs for $E_{u}$ and $E_{g}$ states in graphyne-4 are similar to $E_{g}$ and $E_{u}$ in graphyne-2, respectively,
and it explains the opposite signs of $\tilde{m}_{11}$ for these two cases.

\begin{figure}[b]
\begin{center}
	\includegraphics[width=85mm]{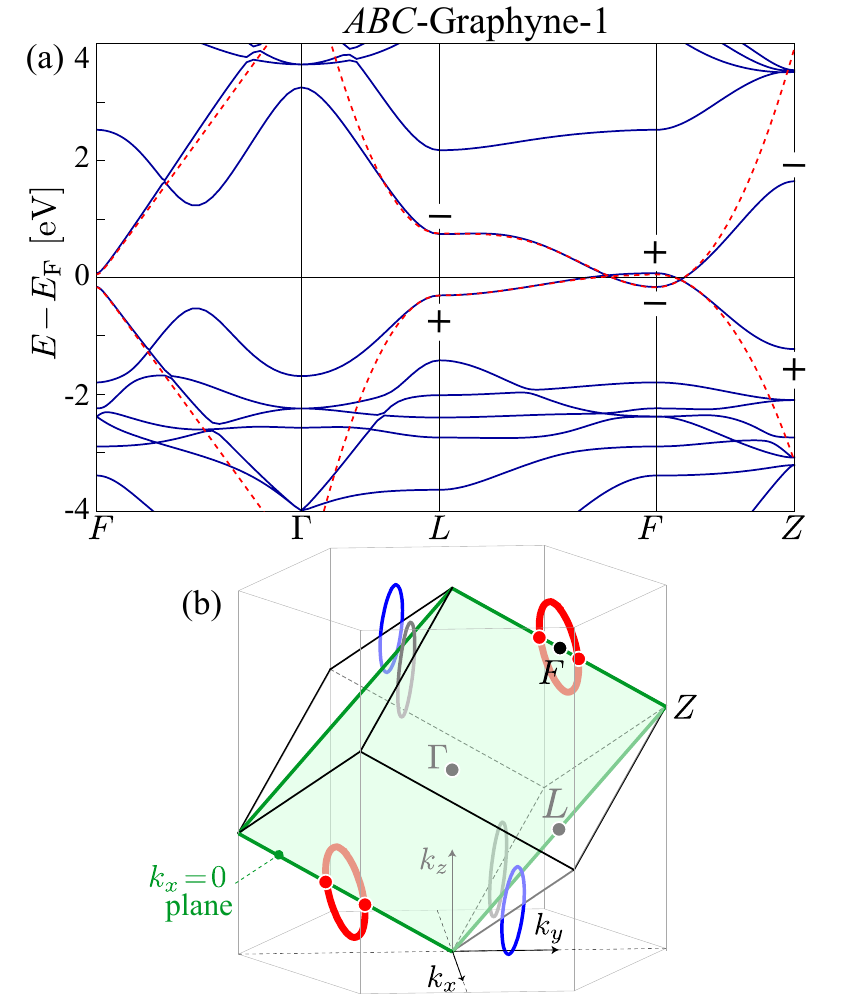}
	\caption{
	(a) Band structure of ABC stacked graphyne-1 obtained from the DFT calculation.
	The dashed line is the dispersion relation calculated by  the effective model Eqs.~(\ref{eq:2x2}) with (\ref{eq:gcond}).
	The sign indicates the mirror parity $\eta_{n}(M_x)=\pm 1$  at high symmetric points.
	(b) The nodal line structure in the three-dimensional momentum space.
	The red, blue and gray loops are $C_{3\bm{z}}$ counterparts.
	}\label{fig:abcband1}
\end{center}
\end{figure}

\begin{table*}[t]
	\caption{Model parameters of Eq.~(\ref{eq:gcond}) for ABC stacked graphyne-1 and 3 to reproduce the DFT results.
	} \label{table:abcodd}
	\renewcommand{\arraystretch}{1.5}
	{\tabcolsep = 4.0mm
	\begin{tabular*}{180mm}{lrrrrrrrrrrr}
		\hline 
		\hline 
		$N$& $E_{00}$ & $E_{01}$ &$E_{02}$& $B_{00}$ & $D_{01}$ &$E_{z0}$ & $E_{z1}$ & $E_{z2}$ & $B_{z0}$ & $D_{z1}$ & $v/L$ \\
		\hline
		1 & 0.15 & $-0.14$ & $-0.06$ & 0.01 & $-0.05$ & $ 0.26$ & $-0.33$ & $-0.05$   & $-0.33$    & $0.3$ & 0.77 \\
		3 & 0.16 & $-0.02$ & $-0.03$ & 0.02 & $-0.07$ & $-0.26$ & $-0.02$ & 0.01   & $-0.16$ & $0.1$  & 0.36 \\
		\hline 
		\hline
	\end{tabular*}
	}
	\renewcommand{\arraystretch}{1}
\end{table*}

\section{ABC-stacked graphynes (Odd $N$)}\label{sec:abcodd}

The electronic structure of ABC-stacked graphyne with odd $N$ is completely different from that of even $N$,
due to the even-odd effect in monolayer graphynes.
In the following, we study the graphyne-1 and -3 paying attention to the band characteristics and the topological properties.

\subsection{ABC stacked graphyne-1}\label{sec:abc1}

As shown in Fig.~\ref{fig:abclat1}, the lattice structure and the crystal symmetry of ABC stacked graphyne-1 are
the same as those of graphyne-2 except for the length of 1D chain in Fig.~\ref{fig:abclat1}(a).
The $M$ point of monolayer Brilloiuin zone, where the gap minima appears in graphyne-1
(see Fig.~\ref{fig:band2d} and ~\ref{fig:tb2d}), is located  
on $FL$ line of the rhombohedral Brillouin zone [Fig.~\ref{fig:abclat1}(b)].

We optimize the atomic structure and calculate electronic structure by  the DFT band calculation 
in a similar manner to the systems with even $N$.
Figure~\ref{fig:abcband1}(a) shows the band structure on the high-symmetry lines depicted in Fig.~\ref{fig:abclat1}(b).
A relatively narrow energy gap around $L$ and $F$ points originates from the minimum gap at $M$ point in the monolayer graphyne-1. 
Importantly, we observe a band crossing on the $LF$ and $FZ$ lines,
which is a cross section of a nodal line protected by mirror reflection symmetry $M_{x}=IC_{2\bm{x}}$.
Since the $k_x=0$ plane is invariant under $M_{x}$, the energy bands with the mirror eigenvalues $\eta({M}_x)=\pm 1$ do not mix with each other on the plane.
In addition, as seen in Fig.~\ref{fig:abcband1}(a), the energy bands of $\eta({M}_x)=\pm 1$ are inverted just near the $F$ point,
resulting in a band-crossing ring around $F$ on the $k_x=0$ plane [Fig.~\ref{fig:abcband1}(b)].
Due to threefold rotational symmetry,  there are three independent nodal rings in the first Brillouin zone.

We derive a two-band effective theory around $F$ point as follows.
The little group at $F$ point in the rhombohedral Brillouin zone
is generated by $C_{2\bm{x}}$ and $I$~\cite{bradlay}. 
As $F$ and $M$ share the same in-plane momentum $(k_x, k_y)$,
the lowest energy states at $F$ originate from $B_{3g}$ and $A_u$ states at $M$ of monolayer graphene-1 [Fig.~\ref{fig:band2d}(a)], 
which are characterized by parity $[\eta(M_x),\eta(I),\eta(C_{2x})]= (+1, +1, -1)$ and $(-1,-1,-1)$, respectively.
We label these lowest energy states in three dimensional system by $\left|\pm ,F \right>$, where the label $\pm$ corresponds to $\eta(M_x)=\pm1$,
respectively.

\begin{table}[b]
	\caption{Model parameters of Eq.~(\ref{eq:gcond}) for monolayer stacked graphyne-1 and 3
	to reproduce the DFT result.} \label{table:monoodd}
	\renewcommand{\arraystretch}{1.5}
	{\tabcolsep = 6.0mm
	\begin{tabular*}{85mm}{ccccc}
		\hline 
		\hline 
		$N$& $E_{00}$ & $E_{z0}$ & $B_{z0}$ & $v/L$ \\
		\hline 
		1& $0.1$ & 0.2  & 0.1 &  0.7\\
		3& $0.26$ & 0.3  & 0.15 &  0.4\\
		\hline 
		\hline
	\end{tabular*}
	}
	\renewcommand{\arraystretch}{1}
\end{table}

The general form of the two-band effective Hamiltonian is given by
\begin{eqnarray}\label{eq:2x2}
	{H}(\tilde{\bm k}) = \epsilon(\tilde{\bm k}){\sigma_0} + \bm{g}(\tilde{\bm k})\cdot {\bm \sigma}, 
\end{eqnarray}
where $\sigma_{\mu=0,x,y,z}$ is unit and Pauli matrices acting on two dimensional basis $\left|\pm,F\right>$, 
$g_{\mu=x,y,z}$ and $\epsilon$ are the real numbers.  
The momentum $\tilde{\bm k}=[(k_x-k_x^{F})L, (k_y-k_y^{F})L, (k_z-k_z^{F})d]$ is a dimensionless wave vector measured 
from the $F$ point, $\bm{k}^F=2\pi [0,(\sqrt{3}L)^{-1}, (3d)^{-1}]$, and 
normalized in units of $(L,L,d)$ for $(k_x,k_y,k_z)$.
In the present system we have time-reversal symmetry given by  
\begin{eqnarray}\label{eq:trs}
	H^\ast(\tilde{\bm k}) = H(-\tilde{\bm k}). 
\end{eqnarray}
In addition, 
the inversion symmetry is
\begin{eqnarray}\label{eq:inv}
	I H(\tilde{\bm k})I^{-1} = H(-\tilde{\bm k})
\end{eqnarray}
with $I=\sigma_z$ because the two basis of the present system has opposite inversion parity. 
The mirror reflection symmetry is 
\begin{eqnarray}\label{eq:mx}
	M_{x}H(\tilde{\bm k})M_x^{-1}\! =\! H(D_{M_x}[\tilde{\bm k}]),
\end{eqnarray}
with $M_{x}\!=\!IC_{2\bm{x}}\!=\!\sigma_z$, 
and $D_{M_{x}}[\tilde{\bm k}]=(-\tilde{k}_x,\tilde{k}_y,\tilde{k}_z)$.
The Eqs.~(\ref{eq:trs}), (\ref{eq:inv}) and (\ref{eq:mx}) yield to the constraints for $\bm{g}$ and $\epsilon$: 
\begin{eqnarray}\label{eq:gcond}
	\begin{array}{c}
	g_x = 0, \\
	g_{y}(\tilde{\bm{k}})=-g_{y}(-\tilde{\bm{k}})=-g_{y}(-\tilde{k}_x,\tilde{k}_y,\tilde{k}_z) \\ 
	g_{z}(\tilde{\bm{k}})=+g_{z}(-\tilde{\bm{k}})=+g_{z}(-\tilde{k}_x,\tilde{k}_y,\tilde{k}_z) \\
	\epsilon(\tilde{\bm{k}})=+\epsilon(-\tilde{\bm{k}})=+\epsilon(-\tilde{k}_x,\tilde{k}_y,\tilde{k}_z)
	\end{array}
\end{eqnarray}
Up to the second order of $\tilde{k}_x$ and $\tilde{k}_y$,
they are uniquely determined as
\begin{eqnarray}\label{eq:ge}
	g_{y} &\!=\!& (v/L) \tilde{k}_x\nn\\ 
	g_{z} &\!=\!& \sum_{q=0} [E_{zq}\! +\! A_{zq}\tilde{k}_x^2\! +\! B_{zq}\tilde{k}_y^2]  \cos (q \tilde{k}_z)  \!+\! \sum_{q=1} D_{z q} \tilde{k}_y \sin (q\tilde{k}_z) \nn\\
	\epsilon &\!=\!& \sum_{q=0} [E_{0q} \!+\! A_{0q}\tilde{k}_x^2 \!+\! B_{0q}\tilde{k}_y^2]  \cos (q \tilde{k}_z)  \!+\! \sum_{q=1} D_{0 q} \tilde{k}_y \sin (q\tilde{k}_z) \nn\\
\end{eqnarray}
where $E_{\mu q}$, $A_{\mu q}$, $B_{\mu q}$, $D_{\mu q}$ ($\mu=0, z$) and $v$ are model parameters.
Table~\ref{table:abcodd} lists the parameters for the ABC-graphyne-1, which are obtained 
to fit the DFT band structure [dashed curves in Fig.~\ref{fig:abcband1}(a)].

Actually, the effective model Eq.~(\ref{eq:2x2}) 
applies to ABC-graphynes with any odd $N$'s since it is derived purely from the symmetry consideration. 
Also, it is worth noting that the effective model neglecting $k_z$-dependent (i.e., $q\ge 1$) terms 
describes the monolayer graphyne of odd $N$. 
Table~\ref{table:monoodd} presents the parameters
for monolayer graphyne-1 and 3, which are obtained to give the dashed curves in Fig.~\ref{fig:band2d}(a) and (c), respectively.

The trajectory of the nodal line can be obtained by $\bm{g}(\tilde{\bm k})=0$.
As is obvious from Eq.\ (\ref{eq:ge}), the condition becomes $k_x=0$ and $g_{z}=0$, which gives red nodal loops in Fig.~\ref{fig:abcband1}(b).
On the symmetric plane $k_x=0$, the vector $\bm{g}$ is oriented to either of $\pm \bm{z}$, 
and the domain of $g_{z}>0$ and that of $g_{z}<0$ are separated by the nodal line.
On a closed path encircling the nodal line, 
the vector $\bm{g}=(0,g_y,g_z)$ rotates on $g_y$-$g_z$ space by an odd number,
considering the constraint of Eq.\ (\ref{eq:gcond}).
Therefore, the Berry phase on the path is $\pi$, which topologically protects the nodal line~\cite{fangreview}.

\subsection{ABC stacked graphyne-3}\label{sec:abc3}

\begin{figure}[b]
\begin{center}
	\includegraphics[width=85mm]{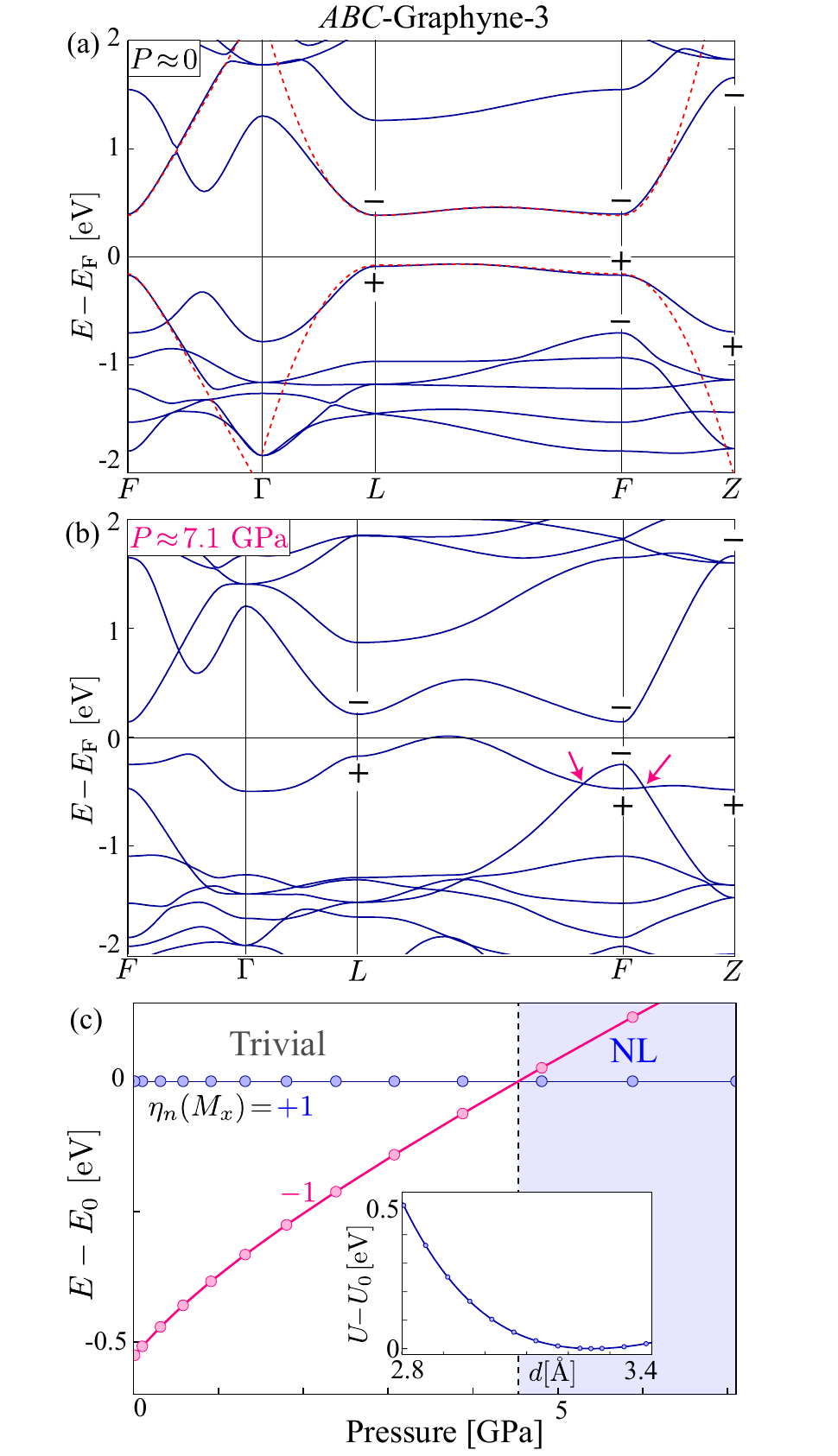}
	\caption{ 
	Energy spectrum of ABC-stacked graphyne-3 obtained from DFT calculation,
	with interlayer distance (a) $d \approx 3.25$ \AA\, ($P\approx 0$) and (b) 2.66 \AA\, ($P\approx 7.1$ GPa).
	The sign on bands indicates the mirror parity $\eta_{n}(M_x)=\pm 1$ at high symmetric points.
	(c) Energy difference between the two highest valence bands at $F$ point as a function of pressure. 
	Inset shows the  total energy of the system (measured from its minimum $U_0$) as a function of the interlayer distance $d$.
	}\label{fig:abc3}
\end{center}
\end{figure}

Finally we discuss the graphyne-3 as an example of the larger odd $N$ system. 
Its low energy effective theory is the same as graphyne-1 and it is given by Eqs.~(\ref{eq:2x2}) and (\ref{eq:ge}).
In Fig.~\ref{fig:abc3}(a), we see that the DFT band structure can be fitted by the continuum model (dashed curves) by using parameters in Table~\ref{table:abcodd}.
In contrast to graphyne-1, the band inversion does not occur, and 
this is because the interlayer atomic overlap in graphyne-3  is smaller than in graphyne-1, 
just in the same manner as the comparison of graphyne-4 to -2.  

Figure \ref{fig:abc3}(b) shows the band structure under pressure of $P = 7.1$GPa.
We see that the band inversion between the valence and conduction bands does not occur 
but instead the two valence bands cross each other.
These two bands are characterized by odd and even mirror parity $\eta(M_x)=-1$ and $+1$, respectively,
and hence the band crossing is formed on a  closed ring around $F$ point, 
similar to the nodal line of graphyne-1 illustrated in Fig~\ref{fig:abcband1}(b). 
The $\mathbb{Z}_2$ monopole charge of this class of nodal line is trivial just as in graphyne-1.
Figure~\ref{fig:abc3}(c) shows the energy difference of the highest two valence bands at $F$ point,
where we see that
the topological phase transition takes place at $P\approx 4.5$ GPa.

\section{Summary}\label{sec:summary}
We have presented a systematic study on the electronic structures and the topological natures of graphyne-$N$ monolayer and ABC-stacked multilayer.
We found an even-odd effect in the $N$-dependence of the band structure,
and in particular, we observed that ABC-stacked graphynes of even and odd $N$'s support two topologically distinct classes of nodal-line semimetal phases.
Specifically, the ABC-graphyne of even $N$ becomes a nodal line semimetal with nontrivial $\mathbb{Z}_2$ monopole charge,
as a consequence of the band inversion of doubly degenerate states in its monolayer counterpart.
The ABC graphyne with odd $N$ only also becomes a nodal line semimetal but without $\mathbb{Z}_2$ monopole charge,
because it is resulting from a band inversion of non-degenerate conduction and valance bands.
ABC graphynes $N=3$ and $4$ become gapped trivial insulators because of smaller interlayer couplings,
while we demonstrate that the external pressure induces the topological phase transitions to nodal-line semimetal phases.
Therefore, graphynes serve as a novel platform to study the physics of $\mathbb{Z}_2$ trivial and non-trivial nodal-line semimetals.

\begin{acknowledgments}
This work was supported by JSPS KAKENHI Grant numbers JP16K17755, JP20K14415, JP17K05496, JP20H01840, and JP20H00127.      
T.K. is partially supported by JSPS Core-to-Core program and  CREST, JST (Grant No. JPMJCR18T4).
A part of numerical calculations was performed on XC40 at YITP in Kyoto University.
\end{acknowledgments}

\appendix 
\section{Computational Details}\label{sec:cmpt}
The DFT calculation in this paper are performed using first principle package {\sc quantum espresso}~\cite{giannozzi2009}.
We employ the ultrasoft pseudopotentials with Perdew-Zunger self-interaction corrected density functional, 
the cutoff energy of the plane-wave basis and the charge density expansion 60 Ry and 300 Ry, respectively, 
and the convergence criterion of 10$^{-8}$ Ry. 
We take the wave number mesh as $12\times12$ for monolayer graphyne $N=$1, 2, 3, and 4. 
The mesh taken for ABC-stacked graphynes are $12\times 12 \times 12$ for $N=1$ and 2, and $4\times4\times8$ for $N=3$ and 4. 
The atomic structure and lattice structure are optimized by the structural relaxation code in {\sc quantum espresso}. 
Here the criterion for the structural relaxation for total energy convergence and force on atoms are taken as $10^{-5}$ Ry
and $10^{-4}$ Ry/$a_{\mathrm{B}}$ with Bohr radius $a_{\mathrm{B}}$, respectively.

\section{The symmetry of effective kagome model} \label{sec:kagome}
Let us consider the electronic property of the effective kagome lattice composed of 1D chain (see Figs.~\ref{fig:chain} and ~\ref{fig:kagome}).
According to the definition of the Bloch state of the present kagome lattice Eq.~(\ref{eq:bloch}), 
the matrix representation of symmetry operator acting on the Bloch Hamiltonian~Eq.~(\ref{eq:hkagome}) is given by
\begin{eqnarray}
	C_{6\bm{z}} &=& 
	\eta_n^{\mathrm{1D}}
	\left(\begin{array}{ccc}
		0 & 1 & 0 \\
		0 & 0 & 1 \\
		1 & 0 & 0
	\end{array}\right), \\
	C_{2\bm{x}} &=& 
	\left(\begin{array}{ccc}
		-1 & 0 & 0 \\
		0 & 0 & -1 \\
		0 & -1 & 0
	\end{array}\right), \\
	I&=& 
	- \eta_{n}^{\mathrm{1D}}, 
\end{eqnarray}
where $\eta_{n}^{\mathrm{1D}}$ is parity of $n$-th 1D chain state defined in Eq.~(\ref{eq:c2parity}).
This implies that the orbital with even (odd) $n$ at each site of kagome lattice can be regarded as $p_z$ ($d_{xz}$) orbital 
at kagome cite with $x$-axis along the direction of 1D chain.

Solving Bloch equation for the Hamiltonian Eq.~(\ref{eq:hkagome}), 
we obtain the wave function at the $\Gamma$ point with $E=\epsilon_n + 2t_{\mathrm{eff}}$, 
\begin{eqnarray}
	\bm{\Psi}_{n,{\pm},0} = \frac{1}{\sqrt{3}}(1,\omega^{\pm1},\omega^{\mp1})^{T}.
\end{eqnarray}
The symmetry characters of this state are given by
$C_{6\bm{z}}\bm{\Psi}_{n,\pm,0} = \eta_{n}(C_{6\bm{z}}) \bm{\Psi}_{n,\pm,0}$
with
\begin{eqnarray}
	\eta_n(C_{6\bm{z}}) = \eta_{n}^{\mathrm{1D}} \omega^{\pm1}
\end{eqnarray}
and by $I\bm{\Psi}_{n,{\pm},\Gamma} = \eta_{n}(I) \bm{\Psi}_{n,{\pm},\Gamma}$ with 
\begin{eqnarray}
	\eta_n(I) = -\eta_{n}^{\mathrm{1D}}.
\end{eqnarray}
These characters for even and odd $n$ correspond to $E_{2u}$ and $E_{1g}$ representations of $D_{6h}$ respectively.

At the $M$ point $(k_x,k_y)=[0,2\pi/(\sqrt{3}L)]=\bm{G}_2/2$ the wave function with eigen energy $E=\epsilon_n -2t_{\mathrm{eff}}$ is 
\begin{eqnarray}
	\bm{\Psi}_{n,\bm{G}_2/2} = \frac{1}{\sqrt{2}}(0,1,1)^{T}.
\end{eqnarray}
This momentum is invariant under $C_{2\bm{x}}$ and $I$ followed by reciprocal lattice translation
$V_{-\bm{G}_2}=\mathrm{diag}[e^{i\bm{G}_2\cdot\bm{L}_1/2},e^{i\bm{G}_2\cdot\bm{L}_2/2},e^{i\bm{G}_2\cdot\bm{L}_3/2}]$, 
which characterize the wave function as
\begin{eqnarray}
	V_{-\bm{G}_2} I \bm{\Psi}_{n,\bm{G}_2/2} = \eta_n(I) \bm{\Psi}_{n,\bm{G}_2/2} \\
	V_{-\bm{G}_2} C_{2\bm{x}}  \bm{\Psi}_{n,\bm{G}_2/2} = \eta_n(C_{2\bm{x}}) \bm{\Psi}_{n,\bm{G}_2/2}.
\end{eqnarray}
The parity for each operation is then identified as
\begin{eqnarray}
	\eta_n(I) = \eta_{n}^{\mathrm{1D}}, \quad \hbox{and} \quad 
	\eta_n(C_{2\bm{x}}) = 1
\end{eqnarray}
These characters for even and odd $n$ correspond to $B_{3g}$ and $A_{u}$ representations of $D_{2h}$ respectively.

\section{Four band effective model of ABC graphynes of even $N$} \label{sec:fourband}
In this section, we derive a low energy effective model for ABC-stacked graphynes of even $N$ 
on the basis of symmetry consideration.
The $\Gamma$ and $Z$ point in ABC-stacked graphyne has the highest symmetry $D_{3d}$ in the momentum space, 
which is generated by improper rotation $S_{6}=IC_{3\bm{z}}$ and twofold rotation $C_{2\bm{x}}$.
At these momenta, in addition, low energy states in ABC-stacked graphynes of even $N$
belong to 2D irreducible representations $E_{g}$ and $E_{u}$ of $D_{3d}$
and these representations mix with each other in the general momenta away from $\Gamma$ and $Z$ point.

We here define the four-dimensional basis as $| \alpha, \beta \rangle$ ($\alpha,\beta = \pm$) and 
unit and Pauli matrix $\tau_\mu$ and $\sigma_\mu$ with $\mu=0, x, y, z$ 
acting on the $2\times 2$ space spanned by first and second index $\alpha$ and $\beta$, respectively. 
More specifically, $| \alpha, \beta \rangle$ are the eigenstates of $\sigma_z$ and $\tau_z$, 
\begin{eqnarray}
	\tau_z | \alpha, \beta \rangle = \alpha |\alpha, \beta \rangle, \quad \hbox{and} \quad 
	\sigma_z | \alpha, \beta \rangle = \beta |\alpha, \beta \rangle.
\end{eqnarray}

Let us consider representations for the generator of $D_{3d}$ group.
Because $E_{g}$ ($E_{u}$) states consist of the eigenstates of 
$S_{6}$ with eigenvalue $e^{\pm i2\pi/3}$ ($-e^{\pm i2\pi/3}$), 
we take a representation with
\begin{eqnarray}\label{eq:S6rep}
	S_{6} = e^{-i 2\pi /3 \sigma_z}\tau_z.
\end{eqnarray}
The representation of twofold rotation satisfying $C_{2\bm{x}} S_{6} C_{2\bm{x}}^{-1} = S_{6}^{-1}$
is then taken as 
\begin{eqnarray}\label{eq:C2rep}
	C_{2\bm{x}} = \sigma_x \tau_0
\end{eqnarray}
We also give the representation of the time-reversal symmetry as
\begin{eqnarray}\label{eq:Trep}
	\mathcal{T} = \sigma_x \tau_z \mathcal{K}
\end{eqnarray}
such that it satisfies $\mathcal{T}^2=\sigma_0\tau_0$, $[S_{6},\mathcal{T}]=0$, $[C_{2\bm{x}},\mathcal{T}]=0$.
Here, $\mathcal{K}$ is complex conjugate operator.
Note that in the present representation, 
first index of basis $\alpha=+$ and $-$ is a label for the $E_{g}$ and $E_{u}$ states at $\Gamma$ and $Z$ point, 
and $\beta=\pm$ is internal degrees of freedom in twofold degeneracy, 
corresponding to the sign of the angular momentum [see Eq.~(\ref{eq:S6rep})].
The representation taken here is not unique and one can use another representation obtained by unitary transformation.

We here consider the effective theory by using this representation.
Around the $k_z$ axis, we can formally expand the Hamiltonian as
\begin{eqnarray}\label{eq:hexp}
	H(\bm{k}) = M(k_z) + \bm{V}(k_z)\cdot \bm{k}_\perp
\end{eqnarray}
with $\bm{k}_\perp=(k_x, k_y)$, $\bm{V}=(V_x, V_y)$. 
$M$ and $V_i$ are $4\times 4$ Hermite matrices, 
which can be expanded as 
\begin{eqnarray}
	X(k_z) = X^{\mu\nu}(k_z) h_{\mu\nu}. 
\end{eqnarray}
with Hermite basis $h_{\mu\nu}=\sigma_\mu \tau_\nu$, real coefficient $X^{\mu\nu}$, and $X=M$, $V_x$ or $V_y$.

Let us narrow down the possible terms of $M(k_z)$ and $\bm{V}(k_z)$ to satisfy the symmetry of the present system,
\begin{eqnarray}
	S_{6}H(D_{S_6^{-1}}[\bm{k}])S_{6}^{-1} = H(\bm{k}), \label{eq:S6} \\
	C_{2\bm{x}}H(D_{C_{2{\bm x}}^{-1}}[\bm{k}])C_{2\bm{x}}^{-1} = H(\bm{k}) \label{eq:C2}, \\
	\mathcal{T}H(-\bm{k})\mathcal{T}^{-1} = H(\bm{k}). \label{eq:T} 
\end{eqnarray}
Here improper rotation $S_6$ of momentum $\bm{k}$ is given by
\begin{eqnarray}
	D_{S_6^{-1}}
	\left[\begin{array}{c}
	k_{x}\\
	k_{y}
	\end{array}\right]
	= D_{C_{6\bm{z}}^{-1}}
	\left[\begin{array}{c}
	k_{x}\\
	k_{y}
	\end{array}\right] = 
	\frac{1}{2}\left(\begin{array}{cc}
	1 & -\sqrt{3}\\
	\sqrt{3} & 1
	\end{array}\right)
	\left(\begin{array}{c}
	k_{x}\\
	k_{y}
	\end{array}\right) \nn
\end{eqnarray}
and 
\begin{eqnarray}
	D_{S_6^{-1}}[k_z] = -k_z
\end{eqnarray}

First, we consider the $\bm{k}$-conserving symmetry obtained from combination of Eqs.(\ref{eq:T}) and (\ref{eq:S6}),
\begin{eqnarray}
	S_{6}^3 \mathcal{T} H(\bm{k}) (S_{6}^3 \mathcal{T})^{-1} = H(\bm{k}) \quad 
\end{eqnarray}
This gives the restriction
\begin{eqnarray}\label{eq:ptrst}
	X^{0y} = X^{xy} = X^{yy} = X^{z0} = X^{zx} = X^{zz} = 0
\end{eqnarray}
for all $X=M$, $V_{x}$, and $V_{y}$. 

\begin{table}[tb]
	\caption{Classification of Hermite basis $h_{\mu\nu} =\sigma_\mu\tau_\nu$ consisting of the $4\times4$ Hamiltonian by $D_{3d}$ point group symmetry. 
	The columns correspond to irreducible representation of $D_{3d}$ point group symmetry, 
	parity for $C_{2\bm{x}}$ rotation, behavior under improper $S_6$ rotation [see Eqs.~(\ref{eq:scalar}) and (\ref{eq:scalar}) for the definition] and the basis matrix.} \label{table:Sclass}
	\renewcommand{\arraystretch}{1.5}
	{\tabcolsep = 2.0mm
	\begin{tabular*}{85mm}{cccc}
		\hline 
		\hline 
		$D_{3d}$ & $C_{2\bm{x}}$ & $S_{6}$ & Basis $h_{\mu\nu}=\sigma_\mu\tau_\nu$ \\
		\hline 
		$A_{1g}$ &$+$& Scalar  & $h_{00}$, $h_{0z}$ \\
		$A_{2g}$ &$-$& Scalar  & $h_{z0}$, $h_{zz}$ \\
		$A_{1u}$ &$+$& Pseudo scalar & $h_{0x}$, $h_{0y}$  \\
		$A_{2u}$ &$-$& Pseudo scalar & $h_{zx}$, $h_{zy}$ \\
		$E_{g}$  &$(-,+)$& Pseudo vector & $(h_{yz},-h_{xz})$, $(h_{y0},-h_{x0})$ \\
		$E_{u}$  &$(+,-)$& Vector  & $(h_{xx},h_{yx})$, $(h_{xy},h_{yy})$ \\
		\hline 
		\hline
	\end{tabular*}
	}
	\renewcommand{\arraystretch}{1}
\end{table}

Next, to consider the symmetry which connects different momentum $\bm{k}$,
we classify the basis $h_{\mu\nu}$ to representations of $D_{3d}$ as in Table~\ref{table:Sclass}, 
using Eqs.~(\ref{eq:S6rep}) and (\ref{eq:C2rep}).
Under $S_6$ operation, each representations behave as either scalar (pseudo scalar)
\begin{eqnarray}\label{eq:scalar}
	S_{6} h_{\mu\nu} S_{6}^{-1} = \pm h_{\mu\nu} , 
\end{eqnarray}
or vector (pseudo vector)
\begin{eqnarray}\label{eq:vector}
	S_{6} 
	\left(\begin{array}{c}
	h_{\mu\nu}\\
	h_{\mu'\nu'}
	\end{array}
	\right) S_{6}^{-1} = 
	\pm \frac{1}{2}\left(\begin{array}{cc}
	1 & -\sqrt{3}\\
	\sqrt{3} & 1
	\end{array}\right)
	\left(\begin{array}{c}
	h_{\mu\nu}\\
	h_{\mu'\nu'}
	\end{array}\right).
\end{eqnarray}
Here upper (lower) signs are for scalar and vector (pseudo scalar and pseudo vector). 
In addition each basis has parity under $C_{2\bm{x}}$ rotation,
\begin{eqnarray}
	C_{2\bm{x}}h_{\mu\nu}C_{2\bm{x}}^{-1}=\pm h_{\mu\nu}.
\end{eqnarray}

The term $M(k_z)$ in Eq.~(\ref{eq:hexp}) independent on the perpendicular momentum $\bm{k}_{\perp}$ 
has restrictions,  
\begin{eqnarray}\label{eq:msym}
	S_{6} M(k_z) S_{6}^{-1} = M(-k_z)\\
	C_{2\bm{x}}M(k_z)C_{2\bm{x}}^{-1} = M(-k_z). 
\end{eqnarray}
according to the Eq.~(\ref{eq:S6}) and (\ref{eq:C2}).
Two types of representations $A_{1g}$ and $A_{2u}$ in Table~\ref{table:Sclass} 
satisfy this condition with coefficients 
$M^{\mu\nu}(-k_z)=M^{\mu\nu}(k_z)$ and $M^{\mu\nu}(-k_z)=-M^{\mu\nu}(k_z)$, respectively. 
Therefore, accompanied with Eq.~(\ref{eq:ptrst}), the possible terms are
\begin{eqnarray}\label{eq:mterm}
	M(k_z) = m_0 h_{00} + m_{1} h_{0z} + m' h_{zy}
\end{eqnarray}
with $m_{i}(-k_z) = m_{i}(k_z)$ and $m'(-k_z) = -m'(k_z)$. 

The inner product $\bm{V}(k_z)\cdot \bm{k}_{\perp}$ in Eq.~(\ref{eq:hexp}) should be scalar of $S_6$ [Eq.~(\ref{eq:S6})]. 
As the momentum $\bm{k}_\perp$ is vector under $S_{6}$ operation,
$\bm{V}$ should be vector (pseudo vector) with coefficient $V^{\mu\nu}(-k_z)=V^{\mu\nu}(k_z)$ [$V^{\mu\nu}(-k_z)=-V^{\mu\nu}(k_z)$].
We see from the Table~\ref{table:Sclass} and Eq.~(\ref{eq:ptrst}) that the possible terms are 
\begin{eqnarray}\label{eq:vterm}
	(V_x,V_y) &=& v'_1 (h_{yz}, - h_{xz}) + v'_2 (h_{y0}, - h_{x0}) \nn \\
	&& + v (h_{xx}, h_{yx}) 
\end{eqnarray}
with $v(-k_z) = v(k_z)$ and $v'_i(-k_z) = -v'_i(k_z)$. 

Finally, summarizing Eq.~(\ref{eq:mterm}) and (\ref{eq:vterm}) and using Pauli matrices $\sigma_\mu$ and $\tau_\nu$, 
the possible form of effective Hamiltonian is
\begin{eqnarray}
	H(\bm{k})\! =\! m_0\! +\! m_1 \sigma_z \!+\! m' \sigma_z\tau_y 
	\!+\! v\tau_x(\sigma_x k_x \!+\! \sigma_y k_y)  \nn\\
	+ (v'_1 \!+\! v'_2 \tau_z ) (\sigma_y k_x \!-\! \sigma_x k_y)
\end{eqnarray}

\bibliography{reference}
\end{document}